\documentclass[pdflatex,sn-mathphys-num]{sn-jnl}

\usepackage{ragged2e}
\usepackage{xcolor}
\usepackage{listings}
\usepackage[bottom]{footmisc}
\usepackage{subfigure}
\usepackage{booktabs}
\usepackage{array}
\usepackage{color}
\usepackage{tikz}
\usepackage{enumitem}

\usepackage{graphicx}%
\usepackage{multirow}%
\usepackage{amsmath,amssymb,amsfonts}%
\usepackage{amsthm}%
\usepackage{mathrsfs}%
\usepackage[title]{appendix}%
\usepackage{textcomp}%
\usepackage{manyfoot}%
\usepackage[most]{tcolorbox}
\usepackage{cleveref}
\Crefname{figure}{Fig.}{Fig.}

\newcolumntype{N}{>{\centering\arraybackslash}m{.5in}}
\newcolumntype{M}{>{\centering\arraybackslash}m{.8in}}
\newcolumntype{G}{>{\centering\arraybackslash}m{2in}}

\newcommand{\aitester}{{\fontfamily{cmss}\selectfont \mbox{AITester}}}

\begingroup
\lccode`\~=`\|
\lowercase{\endgroup
\def\tup#1{{\def~{\;\middle\vert\;}\mathcode`\|="8000\left\langle#1\right\rangle}}}

\colorlet{colexam}{red!60!black}
\tcbset{
  base/.style={
    empty,
    frame engine=path,
    colframe=yellow!6,
    sharp corners,
    title={RQ\thetcbcounter~Result},
    attach boxed title to top left={yshift*=-\tcboxedtitleheight},
    boxed title style={size=minimal, top=4pt, left=4pt},
    coltitle=colexam,fonttitle=\scriptsize\bfseries,
  }
}
\newtcolorbox[use counter=exampleAC]{acrqbox}{%
  base,
  boxed title style={overlay={
    \draw[colexam,line width=3pt,] (frame.north west)--(frame.north east);
  }},
  colback=colexam,
  overlay unbroken={
    \draw[colexam] ([yshift=-1.5pt]title.north east)--([xshift=-0.5pt, yshift=-1.5pt]title.north-|frame.east);
  },
}

\colorlet{colexam}{red!60!black}
\tcbset{
  base/.style={
    empty,
    frame engine=path,
    colframe=yellow!6,
    sharp corners,
    title={RQ\thetcbcounter~Result},
    attach boxed title to top left={yshift*=-\tcboxedtitleheight},
    boxed title style={size=minimal, top=4pt, left=4pt},
    coltitle=colexam,fonttitle=\scriptsize\bfseries,
  }
}
\newtcolorbox[use counter=exampleCDS]{cdsrqbox}{%
  base,
  boxed title style={overlay={
    \draw[colexam,line width=3pt,] (frame.north west)--(frame.north east);
  }},
  colback=colexam,
  overlay unbroken={
    \draw[colexam] ([yshift=-1.5pt]title.north east)--([xshift=-0.5pt, yshift=-1.5pt]title.north-|frame.east);
  },
}

\begin{document}

\newcommand{\dottedline}{\raisebox{2pt}{\tikz{\draw[->,black,dashed,line width = 1pt](0,0) -- (5mm,0);}}}

\newcommand{\blackarrow}{\raisebox{2pt}{\tikz{\draw[->,black,solid,line width = 1pt](0,0) -- (5mm,0);}}}

\title[UAST]{Automated System-level Testing of Unmanned Aerial Systems}

\author*[1]{\fnm{Hassan} \sur{Sartaj}}\email{hassan.sartaj@nu.edu.pk}
\author[2]{\fnm{Asmar} \sur{Muqeet}}\email{asmar.muqeet@questlab.pk}
\author[2]{\fnm{Muhammad Zohaib} \sur{Iqbal}}\email{zohaib.iqbal@questlab.pk}
\author[1]{\fnm{Muhammad Uzair} \sur{Khan}}\email{uzair.khan@nu.edu.pk}

\affil[1]{\orgdiv{Department of Computer Science}, \orgname{National University of Computer and Emerging Sciences}, \orgaddress{\street{A.K. Brohi Rd.}, \city{Islamabad}, \country{Pakistan}}}
\affil[2]{\orgname{Quest Lab}, \orgaddress{\street{I8-Markaz}, \city{Islamabad}, \country{Pakistan}}}

\abstract{
	Unmanned aerial systems (UAS) rely on various avionics systems that are safety-critical and mission-critical. 
	A major requirement of international safety standards is to perform rigorous system-level testing of avionics software systems. 
	The current industrial practice is to manually create test scenarios, manually/automatically execute these scenarios using simulators, and manually evaluate outcomes. 
	The test scenarios typically consist of setting certain flight or environment conditions and testing the system under test in these settings. 
	The state-of-the-art approaches for this purpose also require manual test scenario development and evaluation. 
	In this paper, we propose a novel approach to automate the system-level testing of the UAS. 
	The proposed approach (namely \aitester{}) utilizes model-based testing and artificial intelligence (AI) techniques to automatically generate, execute, and evaluate various test scenarios. 
	The test scenarios are generated on the fly, i.e., during test execution based on the environmental context at runtime. The approach is supported by a toolset. 
	We empirically evaluated the proposed approach on two core components of UAS, an autopilot system of an unmanned aerial vehicle (UAV) and cockpit display systems (CDS) of the ground control station (GCS). 
	The results show that the \aitester{} effectively generates test scenarios causing deviations from the expected behavior of the UAV autopilot and reveals potential flaws in the GCS-CDS.
}


\maketitle

\section{Introduction}\label{sec:introduction}
Unmanned aerial systems (UAS) mainly consist of an unmanned aerial vehicle (UAV) and a ground control station (GCS). 
A UAV is an aircraft that does not have a human pilot on-board. 
It is either operated remotely from the ground control station or autonomously using the autopilot system. 
UAS are becoming progressively important due to their large number of applications in a variety of domains. 
For example, UAS are used for rescue operations, disaster management, geological surveys, agriculture and farming, weather forecasting, wildlife monitoring, entertainment, surveillance, airstrikes, military combat, and explosive material detection. 
UAS rely on several avionics software systems (e.g., an autopilot, navigation system, communication system, etc.) that are critical for completing the mission or are critical to safety. 
Malfunctioning of the UAS avionics software can lead to mission failure that can result in human life and financial loss. 
To ensure the dependability of such systems, international safety standards such as DO-178C~\cite{do178} and others~\cite{marques2012stanag} define several guidelines. 
The automated testing of avionics systems is essentially due to the huge amount of cost and effort required to follow these guidelines. 

UAS are a special type of cyber-physical systems (CPS) and are also considered context-aware systems. 
The system-level testing of UAS requires the simulation of the UAV flight behavior with the help of flight simulators. 
It is typically performed at three levels, (i) software-in-the-loop (SIL) testing, (ii) hardware-in-the-loop (HIL) testing, and (iii) pilot-in-the-loop (PIL) testing also known as field testing. 
In SIL testing, the actual avionics software runs while simulating the flight hardware and environment.
In HIL testing, the original avionics hardware parts are used and the environment is simulated. 
In PIL testing, the UAS is tested in a real environment. 
Testing at each level highly depends on the environmental context, i.e., test scenarios created for SIL testing may not be useful for HIL/PIL testing. 
The main challenge in automating system-level testing of the UAS is the generation and evaluation of environmental context-aware test scenarios.

A traditional method of testing UAS at the system level includes a human tester (pilot) who knows UAV flight behavior and has expertise with UAV flight simulators. 
The tester is provided with the system under test (SUT) and the specifications of the SUT. 
Using this information, the tester manually designs test scenarios. 
For each test scenario, the tester simulates the behavior of a UAV in a particular environment with the help of a flight simulator, and at each flight operation, observes the behavior of the SUT by comparing the expected behavior mentioned in the specifications. 
Testing this way usually requires several days and only a limited set of potential scenarios can be tested. 
The existing approaches available in the literature also require manual test scenario development and manual test evaluation (e.g.,~\cite{jung2007modeling, how2008real, wu2019testing, scanavino2019new}). 
The activity of manual testing is tedious and error-prone due to numerous potential scenarios that need to be tested in the extremely dynamic environment of the UAV.

In this paper, we propose a novel approach to automate the system-level testing of UAS. 
To automate the manual testing of UAS, a human-level control on a flight simulator and a mechanism to model the SUT specifications are required. 
Therefore, our approach (namely \aitester{}) utilizes artificial intelligence (AI) and model-based testing (MBT) techniques to automatically generate, execute, and evaluate various test scenarios. 
We use deep reinforcement learning (an AI technique) to get human-level control on a flight simulator and MBT to model the specifications of SUT. 
The test scenarios are generated on the fly, i.e., during test execution based on the environmental context at runtime. 
For this purpose, we propose a UML profile for modeling the structural aspects of UAS SUT and a UML profile for modeling the flight behavior of the UAV. 
The structural profile is used to model the domain concepts of the UAS SUT and the flight behavior profile is used to model the abstract flight behavior of the UAV. 
Our approach requires the constraints (written in Object Constraint Language (OCL)~\cite{ocl}) specified on the UAS SUT domain model and state invariants on the flight behavior model. 
The UAS SUT domain model, the UAV flight behavior model in the form of a UML state machine, and the expected behavior of the SUT in the form of OCL constraints are used as input to our approach. 
Using the UAV flight behavior information, a deep reinforcement learning algorithm is trained to explore different flight test scenarios while interacting with the environment at runtime. 
The flight information received from the environment is used to determine the current state of the UAV. 
This information is also used to populate an instance model of the UAS SUT domain model. 
The instance model representing the current flight information of the UAS SUT is used to evaluate OCL constraints. 
The \aitester{} gets rewards based on the correct selection of action and the number of OCL constraints violated based on that action. 
During the testing process, the \aitester{} continues to explore different flight states, learns to violate OCL constraints (expected behavior of the SUT) based on actions, and compiles the results for further analysis by avionics testers.

We developed a toolset that implements our proposed approach to support testing automation.
We performed two experiments to evaluate \aitester{} using two different subsystems of the UAS. 
For the first experiment, we used an autopilot of the UAV as a SUT, specifically employing the ArduCopter, a widely-used open-source autopilot system.  
For the second experiment, we used the cockpit display systems (CDS) of the GCS as the SUT, particularly utilizing an industrial case study of GCS-CDS. 
For UAV autopilot, the results indicated that \aitester{} effectively generates test scenarios provoking deviations from the expected behavior of the UAV autopilot.
For GCS-CDS, the results showed that \aitester{} is effective in finding four different types of faults and in exploring diverse test scenarios. 
Overall, the results of both experiments demonstrated that \aitester{} effectively generates test scenarios causing deviations from the expected behavior of the SUT that lead to potential faults. 

This paper is an extended version of the research abstract published in the Doctoral Symposium of the 36th IEEE/ACM International Conference on Automated Software Engineering (ASE)~\cite{sartaj2021automated}. 
In the conference paper, an initial idea of the proposed approach targeting UAVs and a pilot experiment on UAV autopilot was presented. 
The primary contributions of this paper relative to the conference paper are summarized below.
\begin{itemize}

	\item We present a comprehensive modeling methodology targeting the whole UAS. This includes UML profiles for modeling the structural and behavioral aspects of a UAS SUT and modeling guidelines for avionics testers. The modeling methodology presented in this paper covers all UAS subsystems, e.g., GCS. 
	\item We propose a novel approach (\aitester{}) that utilizes MBT and AI to automatically generate, execute, and evaluate various test scenarios based on the environmental context at runtime.
	\item We developed a toolset that implements our proposed approach to support testing automation and to enable further research and development. 
	The toolset is publicly available at an online repository\footnote{https://github.com/hassansartaj/uast-toolset\label{uast-repo}}.
	\item We performed two experiments to evaluate \aitester{} using two different subsystems of the UAS. The first experiment was performed using a UAV autopilot as a SUT, extending the pilot experiment with a new research question. The second experiment was performed using the CDS of GCS as a SUT to evaluate the applicability of the proposed approach to another UAS subsystem.

\end{itemize}

The remaining portion of the paper is organized as follows. 
\Cref{sec:bg} presents the background of unmanned aerial vehicles, model-based testing, reinforcement learning, and deep learning. 
\Cref{sec:app} describes our proposed approach in detail. 
\Cref{sec:tool} provides a discussion on the toolset that implements the proposed approach. 
\Cref{sec:eval} provides the empirical evaluation of the proposed approach.
\Cref{sec:rw} discusses the works related to this paper. 
\Cref{sec:limit} outlines the limitations of the proposed approach. 
Finally, \Cref{sec:con} concludes the paper.

\section{Background}\label{sec:bg}
This section presents a concise background discussion on unmanned aerial systems, model-based testing, reinforcement learning, and deep learning.

\subsection{Unmanned Aerial Systems (UAS)}
Unmanned aerial systems (UAS) primarily consist of an unmanned aerial vehicle (UAV) and ground control station (GCS). 
A UAV is an aircraft without a human pilot onboard. 
It is either operated by a remote pilot in the GCS or using an onboard autopilot system. 
To perform communication between UAV and GCS, a data link layer is used in which the uplink is used to send the command to the UAV, and the downlink is used to receive the flight information from the UAV.
UAVs are classified into nine different categories based on their weight, size, operating altitude, range, and endurance~\cite{sadraey2017unmanned}. 
UAV designs include fixed-wing (e.g., plane), single-copter (e.g., helicopter), multi-copter (e.g., quad-copter), and hybrid (e.g., quad-plane). 
UAVs can carry two types of payloads, (i) dispensable payloads such as medical supplies, food items, or ammunition, and (ii) indispensable payloads such as cameras, sensors, or radars of different kinds. 
UAVs have a large number of applications in different civil and military domains. 
In the civil sector, UAVs are used for rescue operations, disaster management, geological surveys, agriculture and farming, weather forecasting, wildlife monitoring, entertainment, etc.
UAVs are also used for different military operations such as intelligence, surveillance and reconnaissance (ISR), airstrikes, military combat, and explosive material detection.

A GCS is a ground-based cockpit for monitoring and controlling a UAV. 
The GCS consists of the cockpit display systems (CDS), a flight control system, a mission planner, a communication channel, a navigation system, 2D/3D maps, and a payload monitoring and control system. 
The CDS of the GCS provides various types of interfaces with different flight instruments to show the UAV flight information~\cite{sartajtesting}. 
These user interfaces display different types of information received from the different avionics systems of a UAV during the flight.
For example, the navigation system generates information such as latitude, longitude, and altitude using sensors such as gyros, accelerometers, and GPS. 
This information is necessary for a UAV to accurately determine its current position, next waypoint to follow, and navigate on the planned route. 
The flight control system is used to generate control and navigational commands for the UAV. 
The mission planner supports UAV profile management, route planning, path optimization, payload setup, and planning unforeseen situations. 
The communication channel is responsible for supporting all types of communications between a UAV and a GCS.
The navigation system is used to monitor the navigational information (such as latitude, longitude, and altitude) of the UAV. 
The 2D/3D maps support interactive mission planning, mission execution, and observing the state of the UAV during flight. 
The payload monitoring and control system helps an operator in observing the status of the payload and in controlling the payload.

\subsection{Model-based Testing}
Model-based testing (MBT) is a testing technique that uses models of the system under test (SUT) to support automation in various testing activities such as test case generation and test execution. 
For conventional system testing, where it is possible to construct complete models of expected behavior, MBT provides a comprehensive solution for automated test generation, test execution, and test  evaluation~\cite{PetrenkoSM12}. 
The MBT process generally comprises five steps~\cite{utting2010practical}. 
In the first step, the abstract models of the SUT are developed based on the requirements or specifications using any modeling language such as Unified Modeling Langauge (UML)~\cite{uml}. 
The SUT models capture the system's functionality in the form of various UML models (diagrams) such as a class diagram, a state machine, and a sequence diagram.
Different UML models support the automation of several testing activities. 
For example, UML state machines support the generation of test sequences~\cite{sartajtesting}.  
The constraints specified in the requirements are modeled using a constraint specification language, i.e., Object Constraint Language (OCL)~\cite{ocl,khan2019aspectocl}. 
The SUT models developed in UML are augmented with the constraints written in OCL.
The second step is to determine the criteria for test case selection.  
The criteria for selecting test cases can be defined based on requirements (e.g., requirement coverage) or using the models of SUT (e.g., state/transition/transition-pair coverage of the state machine). 
In the third step, the specifications of test cases are defined based on the criteria for test selection. 
In the fourth step, the models of SUT (created in the first step) and test case specifications (defined in the third step) are used for generating test cases. 
In the fifth step, the generated test cases are executed using the test execution platform that is suitable for the SUT. 
In the end, the test verdict is determined by comparing the test execution output with the SUT expected output. 
If the expected output matches the test execution output, the verdict is \emph{pass} and \emph{fail} otherwise.

\begin{figure}[!t]
	\centerline{\includegraphics[width=6.9cm,height=2.4cm]{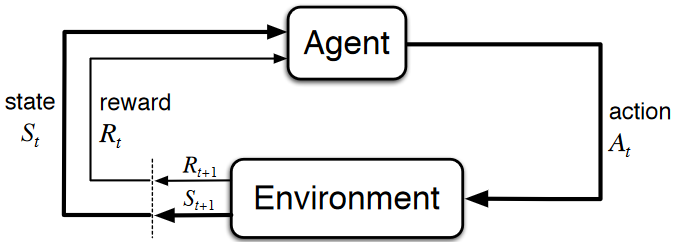}}
	\caption{An agent's interaction with the environment~\cite{sutton2018reinforcement}}
	\label{fig:rl}
\end{figure}

\subsection{Reinforcement Learning (RL)}
Reinforcement learning (RL) is an artificial intelligence technique in which an agent learns based on its interaction with the environment. 
The RL agent, in a particular state, performs some action in the environment. 
Based on the action, the environment returns the next state and the reward.
Therefore, the RL problem can be formulated as a Markov Decision Process (MDP). 
\Cref{fig:rl} shows an MDP in which an agent in a state $S_t$ with reward $R_t$ performs an action $A_t$ on the environment at time \emph{t}.  
Based on the action, the environment returns the next state $S_{t+1}$ and the reward $R_{t+1}$.  
According to the reward received from the environment, the agent learns whether the action was right or wrong. 
The learning process continues until the agent learns an optimal policy that can maximize the expected reward to explore the environment. 
RL is used for multiple applications in a variety of different domains~\cite{li2019reinforcement} such as robotics, transportation, and games. 
In the context of UAVs, it has been used for various purposes such as for the UAV attitude control~\cite{koch2019reinforcement} and the robust control of the autonomous helicopter~\cite{bagnell2001autonomous}.

\subsection{Deep Learning}
It is an AI technique that simulates a part of the human brain. 
For this purpose, it uses artificial neural networks (ANNs) inspired by the biological neural networks of the human brain~\cite{Aggarwal2018}. 
The ANN can be a single-layer network with only input and output layers. 
It can be a multi-layer network with an input layer, an output layer, and several hidden layers in the middle. 
Multi-layer neural networks are also called deep neural networks.
There are three commonly used types of ANNs, (i) feed-forward neural network, (ii) convolutional neural network (CNN), and (iii) recurrent neural network (RNN). 
In a feed-forward neural network, the information from the input layer flows toward the output layer after performing the function approximations in the intermediate levels. 
The feed-forward neural networks are also known as deep feed-forward networks or multi-layer perceptrons. 
In RNN, the information from the input layer flows toward the output layer and can be fed back to the neurons in the previous layer. 
In CNN, the information from the input layer flows toward the output layer by performing different convolution operations in the middle. 

\textcolor{black}{\textbf{Long Short-Term Memory (LSTM).}}
\textcolor{black}{
An LSTM network is a variant of RNNs, specifically designed to handle temporal data and capture the state context. 
It comprises one or more units, each containing input gates, output gates, forget gates, and memory cells. 
The core purpose of the memory cell is to capture the correlation between the current data at time \emph{t} with preceding data points, i.e., \emph{(t-1)!}. 
This mechanism allows LSTM networks to learn longer sequences, making them especially useful for tasks involving sequential data such as time-series analysis. 
}

\section{Approach}\label{sec:app}
In this section, we discuss the proposed approach for the automated system-level testing of the UAS. 
The proposed approach is developed based on the idea presented in previous work~\cite{sartaj2021automated}. 
First, we give a brief overview of the proposed approach. 
Second, we describe the modeling methodology including a discussion of the proposed profiles. 
In the end, we discuss the deep reinforcement learning technique employed in our approach for testing a UAS SUT based on environmental context at runtime.

\begin{figure*}[!t]
    \centerline{\includegraphics[width=\linewidth,height=\textheight,keepaspectratio]{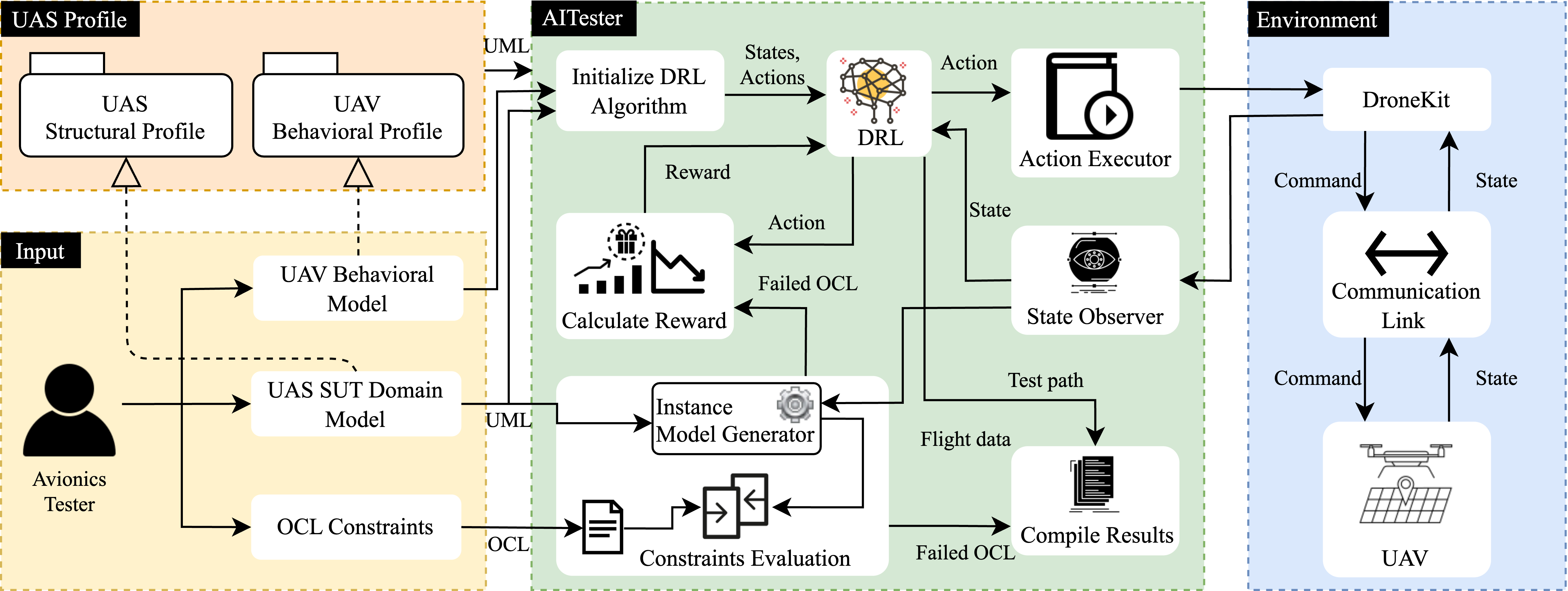}}
	\caption{An overview of the approach showing the proposed UAS profile, inputs required from avionics testers, \aitester{}, and UAV Environment. The dotted arrows (\protect\dottedline) represent the compliance of models to the UAS profile and solid arrows (\protect\blackarrow) represent the information/control flow.}
	\label{fig:ate-app}
\end{figure*}

\subsection{Overview of the Approach}
\Cref{fig:ate-app} shows the overall workflow of the proposed approach. 
To model the system under test, we propose two UML profiles, the UAS structural profile and the UAV behavioral profile. 
The avionics tester is required to model the domain of the UAS SUT using the structural profile and the flight behavior of the UAV using the behavioral profile. 
The approach also requires the expected behavior of UAS SUT in the form of OCL constraints and state invariants. 
The proposed profiles play an important role in supporting the testing automation of the UAS SUT. 
The UAS structural profile represents an observation space of the environment. 
Similarly, the UAV behavioral profile represents the state and action space of the deep reinforcement learning (DRL) algorithm. 
The UAS SUT domain model serves two purposes, (i) initializing the observation space of the environment, and (ii) generating an instance model that will be used to evaluate the OCL constraints during execution. 
The flight behavior model of the UAV is used to initialize the DRL algorithm for UAS SUT. 

Using the UAV flight states and actions, the DRL algorithm starts exploring various flight states using different actions (e.g., takeoff or land). 
The DRL algorithm selects an action and executes it on the environment using DroneKit API. 
The function of DroneKit is to receive the action command for the UAV, send the command to the UAV using a communication link (using Mavlink), and receive the state of the UAV. 
The continuous change in the flight state of the UAV is monitored by a state observer. 
This helps the DRL to keep track of the current state, action, and the next state of the UAV. 
Moreover, the flight data observed during execution is used to populate an instance model of the UAS SUT domain model. 
The instance model is evaluated on the OCL constraints and state invariants. 
The primary objective of our approach is to violate a maximal number of OCL constraints. 
The reward is calculated based on the correct selection of action and the number of OCL constraints violated based on that action. 
In this way, the DRL algorithm continues to explore different flight states, learns to violate OCL constraints (expected behavior of the SUT) based on actions, and compiles the results for further analysis by avionics testers. 
The compiled results include information related to UAV flight states, actions, and the number of failed OCL constraints for each execution.

\subsection{Modeling Methodology for UAS}
A fundamental step in our approach is the modeling of the UAS SUT. 
This includes modeling the UAS SUT domain concepts, the flight behavior of a UAV, and constraints on these models. 
For this purpose, we propose a UAS structural profile for modeling the UAS SUT domain concepts and a UAV behavioral profile for modeling the flight behavior of the UAV. 
The goal of the proposed profiles is to facilitate avionics testers with the modeling of different aspects of UAS SUT using domain-specific terminologies. 
The domains-specific modeling languages available in the literature are either generic for all CPS (e.g.,~\cite{vierhauser2023amon}) or context-specific~\cite{araujo2022testing} (e.g., targeting UAV Swarm~\cite{aloui2022new}). 
Our proposed profiles are based on UML which is widely adopted by the avionics industry~\cite{araujo2022testing}. 
Moreover, the proposed profiles consist of UAS domain concepts that avionics testers are familiar with.
We provide modeling guidelines for avionics testers in the process of modeling a UAV using the proposed profiles and specifying constraints.

\subsubsection{UAS Structural Profile}
\begin{figure}[!t]
	\centerline{\includegraphics[width=\linewidth,height=\textheight,keepaspectratio]{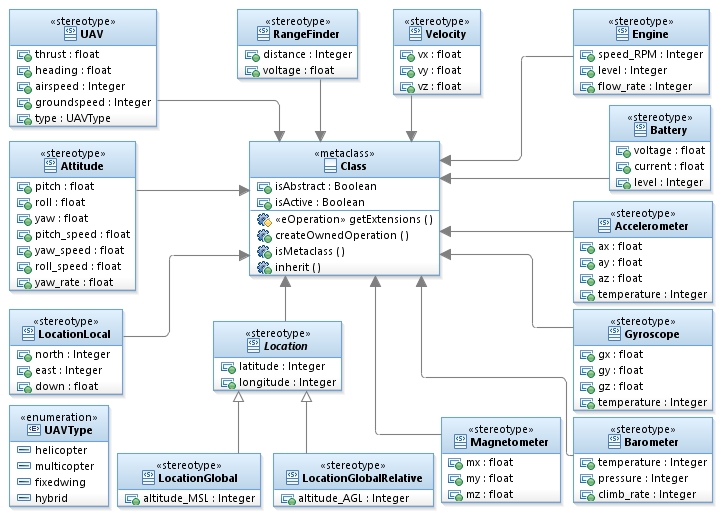}}
	\caption{UAS profile for modeling structural details}
	\label{fig:uav-profile}
\end{figure}

\Cref{fig:uav-profile} shows the profile to model the structural aspects of the UAS SUT. 
A key concept of the profile is \emph{\mbox{UAV}} with properties such as \emph{thrust}, \emph{heading}, \emph{airspeed}, \emph{groundspeed}, and \emph{type}. 
The type of UAV can be a helicopter, multi-copter, fixed-wing, or hybrid UAV. 
The enumeration \emph{\mbox{UAVType}} is used to model various types of UAVs. 
Another important concept is the \emph{\mbox{Attitude}} that represents \emph{pitch}, \emph{roll}, \emph{yaw}, \emph{pitch\_speed}, \emph{roll\_speed}, \emph{yaw\_speed}, and \emph{yaw\_rate} of the UAV. 
The location of a UAV can be broadly categorized into \emph{\mbox{LocationLocal}} and \emph{\mbox{LocationGlobal}}. 
The \emph{\mbox{LocationLocal}} can be represented in degrees using north/south, east/west, and up/down values. 
Whereas the \emph{\mbox{LocationGlobal}} can be represented using \emph{latitude}, \emph{longitude}, and \emph{altitude} values. 
The \emph{altitude} can be measured above mean sea level (i.e., \emph{altitude\_MSL}) and it can be relative to ground level (i.e., \emph{altitude\_AGL}).
The UAV can have a \emph{\mbox{RangeFinder}} that is used to calculate the distance. 
The UAV can have \emph{Velocity} along three axes i.e., \emph{vx}, \emph{vy}, and \emph{vz}. 
To model different power systems, the profile provides two concepts, \emph{\mbox{Engine}} and \emph{\mbox{Battery}}. 
The \emph{Engine} has properties such as \emph{speed}, \emph{level}, and \emph{flow\_rate}. 
The \emph{\mbox{Battery}} has properties such as \emph{voltage}, \emph{current}, and \emph{level}. 
Besides the above-mentioned concepts, the UAV can also have concepts for modeling different measurement units such as \emph{\mbox{Accelerometer}}, \emph{\mbox{Gyroscope}}, \emph{\mbox{Barometer}}, and \emph{\mbox{Magnetometer}}.

\begin{figure}[!t]
	\centerline{\includegraphics[width=\linewidth,height=\textheight,keepaspectratio]{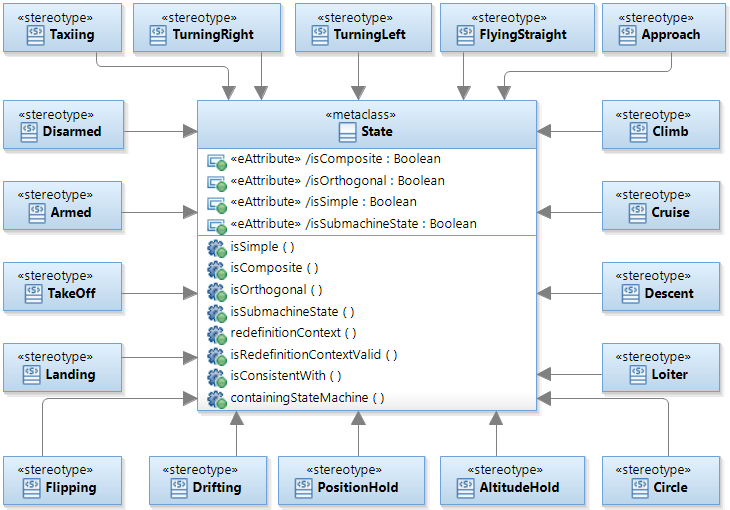}}
	\caption{An excerpt of UAV behavioral profile for modeling flight states}
	\label{fig:uav-beh-profile-1}
\end{figure}

\begin{figure}[!t]
	\centerline{\includegraphics[width=12.2cm,height=6.9cm,keepaspectratio]{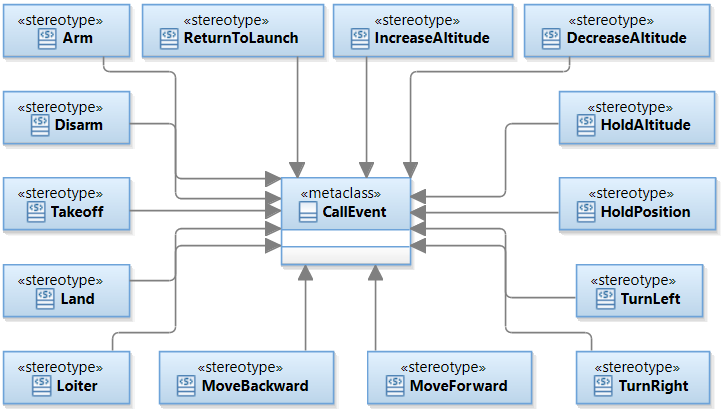}}
	\caption{An excerpt of UAV behavioral profile for modeling flight actions}
	\label{fig:uav-beh-profile-2}
\end{figure}

\subsubsection{UAV Behavioral Profile}
To model the flight behavior of the UAV, the behavioral profile has two parts. 
One for modeling the abstract flight states as shown in \Cref{fig:uav-beh-profile-1}. 
The second part is used for modeling the events that can lead the UAV into various flight states. 
An excerpt of the behavioral profile for modeling events is shown in \Cref{fig:uav-beh-profile-2}.  

The abstract flight states represent various flight stages of the UAV. 
Initially, the UAV is in a \emph{\mbox{Disarmed}} state. 
After arming the motors, the UAV goes to the \emph{\mbox{Armed}} state. 
From the \emph{\mbox{Armed}} state, the UAV can go to the \emph{\mbox{Taxiing}} or \emph{\mbox{TakeOff}} state. 
If the UAV is fixed-wing, it has to start taxi on the ground before it can take off. 
Whereas the single or multi-copter type of UAV can directly take off from the ground without taxiing. 
In the case of hybrid UAVs, both states can be used alternatively. 
After taking off from the ground, the UAV can go through different flight phases such as \emph{\mbox{Climb}}, \emph{\mbox{Cruise}}, \emph{\mbox{Descent}}, \emph{\mbox{PositionHold}}, and  \emph{\mbox{AltitudeHold}}. 
The UAV can perform special types of flights such as \emph{\mbox{Flipping}}, \emph{\mbox{Drifting}}, \emph{\mbox{Loiter}}, and \emph{\mbox{Circle}}. 
The UAV can fly straight when in \emph{\mbox{FlyingStraight}} state, it can take a left turn when in \emph{\mbox{TurningLeft}} state, and it can take a right turn when in \emph{\mbox{TurningRight}} state. 
At the end of the flight, first, the UAV goes to the \emph{\mbox{Approach}} state, and then it goes to the \emph{\mbox{Landing}} state. 

Each abstract flight state is achieved using different events. 
To arm the motors, the \emph{\mbox{Arm}} event is used. 
Similarly, the \emph{\mbox{Disarm}} is used to disarm the motors. 
The event \emph{\mbox{Takeoff}} is used to take off the UAV from the ground and the event \emph{\mbox{Land}} is used to land the UAV on the ground. 
The event \emph{\mbox{ReturnToLaunch}} is used to make the UAV go to the home position (from where it took off) and land at that position. 
During the flight, UAV can move forward or backward (not in the case of fixed-wing) using the events \emph{\mbox{MoveForward}} and \emph{\mbox{MoveBackward}} respectively. 
Similarly, it can make a left or right turn using the events \emph{\mbox{TurnLeft}} and \emph{\mbox{TurnRight}} respectively. 
In the case of multi-copter or hybrid UAV, the event \emph{\mbox{HoldPosition}} makes it maintain the current position. 
The event \emph{\mbox{HoldAltitude}} is used to maintain a stable altitude. 
Whereas the events \emph{\mbox{IncreaseAltitude}} and \emph{\mbox{DecreaseAltitude}} are used to increase or decrease the altitude respectively. 
Note that the profile shown in \Cref{fig:uav-beh-profile-2} has only a subset of all the events modeled in the complete profile. 
The complete profile is available at an online repository\footnote{https://github.com/hassansartaj/uas-profile}.

\subsubsection{Modeling Guidelines}
To model the domain of the UAS SUT, all avionics components (e.g., sensors) need to be modeled as a UML stereotyped class with the stereotype from the UAS structural profile. 
The profile stereotypes contain some common properties of various avionics components. 
The matching properties need to be used as it is and the additional properties can be added to the domain class. 
The avionics tester can freely model the associations among the domain classes.

The flight behavior model of the UAV needs to be modeled using the UML state machine. 
For this purpose, all flight states and events need to be modeled using the states and events stereotypes from the UAV behavioral profile. 
The avionics tester is required to model the end-to-end flight behavior of the UAV, i.e., to take off from the ground, perform different flight operations, and land on the ground. 
For example, a UAV cannot fly in the air without initial ground operations, therefore, the UAV flight state machine without ground states/events will not be useful. 
All the state machine events are required to be modeled as UML \emph{\mbox{CallEvents}}. 
Other types of events such as \emph{\mbox{SignalEvent}}, \emph{\mbox{TimeEvent}}, and \emph{\mbox{ChangeEvent}} are simulator-specific events that can not be controlled using the DRL algorithm. 
In the case of guard conditions, our approach requires modeling the guards in the form of OCL constraints or invariants. 
While modeling transitions, the avionics tester needs to model all possible transitions and events for a flight state.

\subsection{Modeling of UAS SUT}\label{sec:models}
Our approach requires three inputs. 
First, the domain model of the UAS SUT needs to be modeled using the structural profile. 
Second, the flight behavior is in the form of a UML state machine that is modeled using the behavioral profile. 
The third input is the expected behavior of UAS SUT in the form of OCL constraints that include expected ranges of values on various properties of the UAS SUT and w.r.t. the flight states. 
In the following, we discuss the modeling of the UAS SUT domain, flight behavior, and OCL constraints. 

\begin{figure}[!t]
	\centerline{\includegraphics[width=11.3cm,height=6.5cm,keepaspectratio]{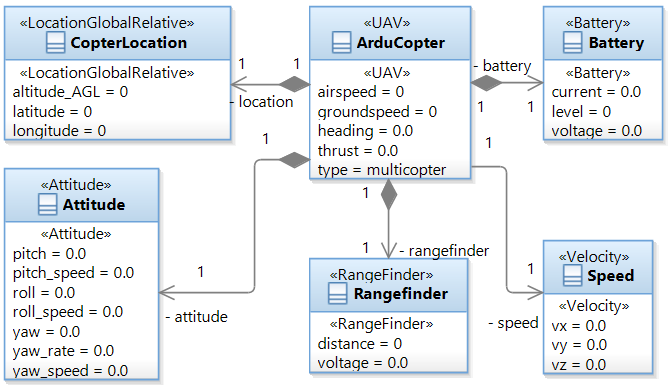}}
	\caption{ArduCopter domain model}
	\label{fig:uav-dm}
\end{figure}

\begin{figure*}[!t]
	\centerline{\includegraphics[width=\linewidth,height=\textheight,keepaspectratio]{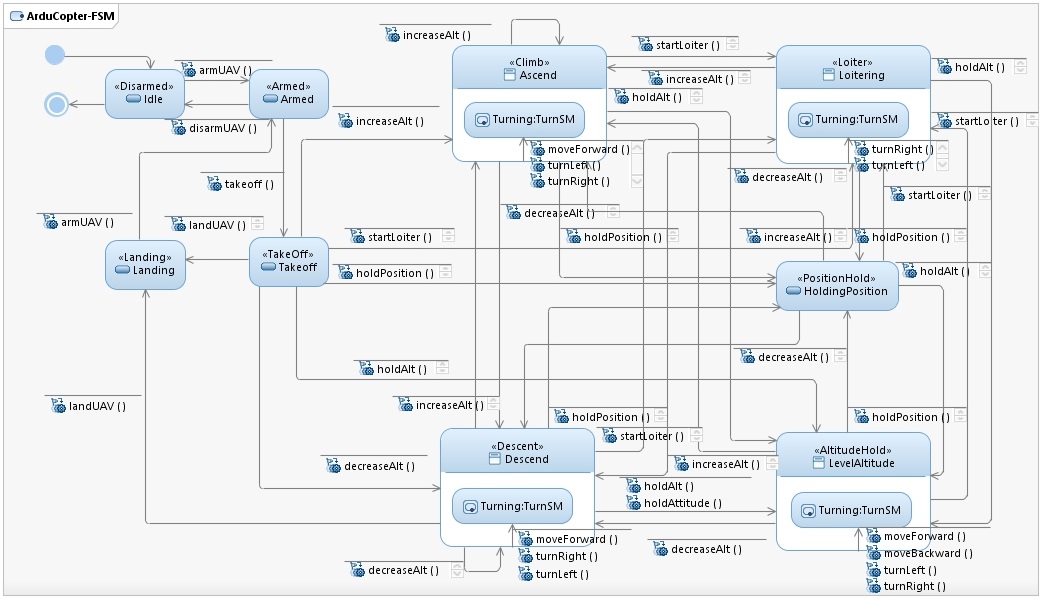}} 
	\caption{ArduCopter flight behavioral model in the form of UML state machine}
	\label{fig:uav-fsm}
\end{figure*}

\begin{figure}[!t]
	\centerline{\includegraphics[width=10.7cm,height=7.4cm,keepaspectratio]{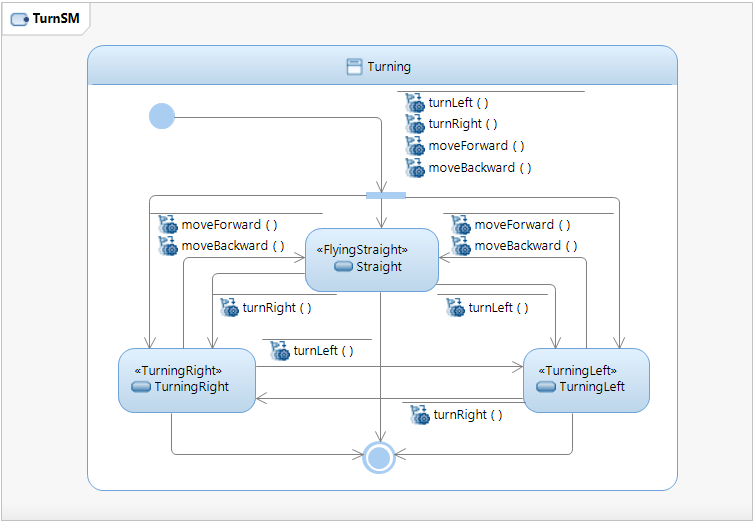}}
	\caption{ArduCopter sub-state machine for turns}
	\label{fig:uav-tsm}
\end{figure}

\subsubsection{UAS SUT Domain Modeling}\label{sbsec:dm}
To model the domain of UAS SUT, an instance of the structural profile is created for the target UAS component. 
\Cref{fig:uav-dm} shows a domain model for the autopilot of a multi-copter named ArduCopter. 
The \emph{\mbox{ArduCopter}} is modeled as an instance of \emph{\mbox{UAV}} meta-class. 
The location of the ArduCopter is measured according to the global relative point, so \emph{\mbox{CopterLocation}} is modeled as an instance of the \emph{\mbox{LocationGlobalRelative}} meta-class. 
The attitude (roll, pitch, and yaw) of the ArduCopter is modeled using the \emph{\mbox{Attitude}} stereotype. 
The ArduCopter has a range finder to calculate the distance which is modeled using the \emph{\mbox{RangeFinder}} stereotype. 
Similarly, the speed of the ArduCopter along the three axes is modeled using the \emph{\mbox{Velocity}} stereotype. 
Finally, the power system used in the ArduCopter is modeled using the \emph{\mbox{Battery}} stereotype. 

The domain model of the UAS SUT serves two purposes. 
First, it is used to provide the UAS SUT domain classes and properties to the environment utilized for the reinforcement learning algorithm (\Cref{sec:drl}). 
This information becomes the observation space of the environment that is observed during the UAV flight. 
The second use of the domain model is to evaluate OCL constraints. 
For this purpose, an instance of the UAS SUT domain model is created and the slots are filled using the UAV flight data. 
The OCL constraints modeled for the corresponding flight state are evaluated on the instance model. 
The evaluation results are used later for the reward calculation (\Cref{sbsec:reward}).

\subsubsection{Flight Behavior Modeling}\label{sbsec:fbm}
The flight behavior of the UAV needs to be represented using abstract flight states and events. 
\Cref{fig:uav-fsm} and \Cref{fig:uav-tsm} show the flight behavior modeled as a UML state machine for the ArduCopter. 
The flight state machine contains various abstract flight states modeled as instances of meta-states from the UAV behavioral profile. 
Similarly, all the events are modeled as instances of various flight events from the UAV behavioral profile. 

Initially, the ArduCopter is in \emph{\mbox{Idle}} state which is modeled using the \emph{\mbox{Disarmed}} profile stereotype. 
To start the motors (arm the UAV) in the \emph{\mbox{Idle}} state, the event \emph{armUAV()} is used to take the ArduCopter to the \emph{\mbox{Armed}} state. 
When the ArduCopter is in the \emph{\mbox{Armed}} state, it can either turn off the motor to go back to the \emph{\mbox{Idle}} state or it can go to \emph{\mbox{Takeoff}} state using the \emph{takeoff()} event. 
After taking off from the ground, the ArduCopter has six options.
It can go to the \emph{\mbox{Ascend}} state using the event \emph{increaseAlt()}, or it can go to the \emph{\mbox{Descend}} state using the event \emph{decreaseAlt()}, or it can go to the \emph{\mbox{Loiter}} state using the event \emph{startLoiter()}, or it can go to the \emph{\mbox{PositionHold}} state using the event \emph{holdPosition()}, or it can go to the \emph{\mbox{AltitudeHold}} state using the event \emph{holdAlt()}, or it can go to the \emph{\mbox{Landing}} state using the event \emph{landUAV()}. 
If its next state is \emph{\mbox{Ascend}}, it can remain in that state, or it can go to the \emph{\mbox{Loiter}} or \emph{\mbox{PositionHold}} or \emph{\mbox{AltitudeHold}} or \emph{\mbox{Descend}}. 
Moreover, the \emph{\mbox{Ascend}} state has a sub-state machine for different types of turns, \emph{Straight}, \emph{\mbox{TurningRight}}, and \emph{\mbox{TurningLeft}} as shown in \Cref{fig:uav-tsm}. 
That means, the ArduCopter can \emph{\mbox{Ascend}} while moving straight or it can take a left turn or right turn.

\emph{\mbox{Ascend}} is the state in which the ArduCopter is making a climb i.e., increasing the altitude. 
\emph{\mbox{Descend}} is the state in which the ArduCopter is making a descending i.e., decreasing the altitude. 
The \emph{\mbox{Descend}} state also has a sub-state machine for turns because the ArduCopter can move straight or it can take a left turn or right turn.
The \emph{\mbox{Loiter}} state represents a flight phase in which the ArduCopter is flying in a small circle. 
During the \emph{\mbox{Loiter}} state, the ArduCopter can loiter in a clockwise or counterclockwise direction. 
Thus, it has a sub-state machine for right and left turns for loitering in a clockwise or counterclockwise direction. 
The state \emph{\mbox{PositionHold}} corresponds to the flight phase in which the ArduCopter is holding its position while keeping the attitude constant. 
Therefore, \emph{\mbox{PositionHold}} state does not have a sub-state machine for different types of turns. 
Similarly, the \emph{\mbox{AltitudeHold}} state represents a flight phase in which the ArduCopter is flying at the same level while keeping the altitude constant. 
For the ArduCopter types of UAV, it is possible to maintain constant altitude and move forward or backward or turn left or right.

The UAV flight behavior modeled in the form of a UML state machine is used by the reinforcement learning algorithm (\Cref{sec:drl}). 
Each abstract flight state becomes a part of the state tuple. 
Each event modeled in the state machine becomes the action space. 
Similarly, all the transitions of the state machine represent the transitions model that will be used to keep track of the current and the next states. 
The UAV state machine plays an integral part in executing UAV flight behavior (i.e., transitioning the UAV through various flight states), which is performed by \aitester{} during the testing process (\Cref{sec:aitester}). 

\subsubsection{Specifying Expected Behavior of the SUT}\label{sbsec:moc}
After modeling the domain UAS SUT and the flight behavior of the UAV, the expected behavior of UAS SUT in the form of OCL constraints is required to be specified. 
The OCL constraints should contain the expected ranges of value on various properties of the UAS SUT and w.r.t. the flight states.
The purpose of the constraints is to check the invalid property values during the UAV flight. 
When formulating OCL constraints corresponding to UAV flight states, the specified ranges for a particular state must be consistent and non-overlapping with the subsequent state. 
For example, if the maximum altitude limit for \emph{\mbox{Takeoff}} state is 50 feet, the altitude range for the subsequent state (\emph{\mbox{Ascend}}) should consider 50 feet as the lower bound. 
Furthermore, the ranges defined in OCL constraints associated with UAV flight states should align with those specified in the general constraints. 

\begin{lstlisting}[label=lst:constraints, language=ocl, caption={OCL constraints including the general and the state invariants}, linewidth=13cm, numbers=none]
--altitude values range (in meters)
C1: context UAV inv: self.location.altitude_AGL>0 and self.location.altitude_AGL<=300
--distance values range (in meters)
C2: context UAV inv: self.rangefinder.distance>0 and self.rangefinder.distance<=5000
--State invariants for Takeoff
C3: context UAV inv: self.oclIsInState(Takeoff) and self.thrust>0 and self.thrust<1
C4: context UAV inv: self.oclIsInState(Takeoff) and self.location.altitude_AGL>0 and self.location.altitude_AGL<=50
--State invariants for Landing
C5: context UAV inv: self.oclIsInState(Landing) and self.airspeed>0 and self.airspeed<=5
C6: context UAV inv: self.oclIsInState(Landing) and self.speed.vz>0 and self.speed.vz<=5
--State invariants for Ascend
C7: context UAV inv: self.oclIsInState(Ascend) and self.airspeed>10 and self.airspeed<100
C8: context UAV inv: self.oclIsInState(Ascend) and self.groundspeed>0 and self.groundspeed<10
--State invariants for Descent
C9: context UAV inv: self.oclIsInState(Descent) and self.airspeed>=5 and self.airspeed<100
C10: context UAV inv: self.oclIsInState(Descent) and self.location.altitude_AGL>10 and self.location.altitude_AGL<100
\end{lstlisting}

\textcolor{black}{
OCL constraints can be derived from two key sources: (i) the UAV configuration manual, and (ii) domain experts or testers who define constraints based on specific testing contexts. 
UAV configuration manuals usually specify parameter ranges for general scenarios and specific flight modes, such as \emph{Takeoff} and \emph{Landing}. 
While the parameter ranges provided by UAV configuration manuals are useful, they typically represent general or broader scenarios. 
Testers often focus on evaluating UAV application-specific or context-dependent conditions. 
For example, a tester might need to assess parameters like the UAV's flight range, position, and navigation in a crop monitoring context. 
Similarly, in a package delivery scenario, a tester could be interested in evaluating parameters such as maximum weight, speed, and battery life along the intended flight path. 
For testing such conditions, testers must define OCL constraints considering each testing scenario. 
In our approach, these constraints are utilized to analyze deviations from the expected behavior of the SUT without imposing any restrictions on UAV flight operations.
Therefore, OCL constraints must be carefully modeled to capture the testable range of values for the various properties during different flight states. 
It is important to note that each UAV has its unique set of constraints. 
The constraints defined for one specific UAV may not necessarily apply to other models of UAVs. 
}

\textcolor{black}{ 
\Cref{lst:constraints} presents a set of example constraints designed to demonstrate the process of constraint specification, without being specific to any UAV type. 
Constraints \emph{C1} and \emph{C2} define general value ranges for \textit{altitude\_AGL} and \textit{distance} properties of a UAV. 
These constraints include ranges that could be obtained from a UAV's configuration manual. 
Constraint \emph{C1} specifies that a UAV can reach up to a maximum altitude of 300m above ground level (AGL). 
Exceeding this limit indicates a deviation from the expected behavior, suggesting a potential malfunction in the SUT. 
Similarly, constraint \emph{C2} defines a range for distance measurement by the UAV's rangefinder, which is particularly beneficial for avoiding potential obstacles. 
Constraints \emph{C3} and \emph{C4} specify the valid ranges for the \emph{thrust} and 
\emph{altitude\_AGL} during the \emph{\mbox{Takeoff}} state. 
In the same way, constraints \emph{C5} and \emph{C6} define the valid range of values for the \emph{airspeed} and 
\emph{speed} along the z-axis during the \emph{\mbox{Landing}} state. 
The constraints for the \emph{\mbox{Takeoff}} and \emph{\mbox{Landing}} states can be derived from the UAV's configuration manual, as it defines the normal ranges for critical flight modes. 
The takeoff state is crucial for initiating a stable flight for the UAV, while the landing state is essential for the UAV's safe return to the ground. 
For instance, a UAV surpassing the takeoff altitude limit could lose control, and violating landing speeds may risk a crash. 
Constraints \emph{C7} and \emph{C8} define ranges for \emph{airspeed} (i.e., between 10 and 100 knots) and \emph{groundspeed} (i.e., between 0 and 10 knots) during the \emph{\mbox{Ascend}} state. 
With these constraints, a tester would be interested in examining the safety scenarios related to the UAV's airspeed and groundspeed during the ascend/climb phase.
For example, maintaining a low airspeed might risk stalling by compromising the UAV's lift and a high groundspeed could destabilize the UAV. 
Constraints \emph{C9} and \emph{C10} specify valid ranges for \emph{airspeed} (specifically between 5 and 100 knots) and \emph{altitude\_AGL} (specifically between 10 and 100 meter) during the \emph{\mbox{Descend}} state. 
These constraints are necessary for UAV's descending phase and a tester might be interested in testing scenarios related to this phase. 
For instance, a high airspeed or a low altitude could potentially result in a UAV crash.
}

\subsection{System-level Testing of UAS}\label{sec:drl}
A typical practice of the avionics tester is to simulate the flight behavior of the UAV using the SIL/HIL simulator.  
During the simulation, the tester performs a certain action, observes the next state of the UAV in an environment, and analyzes the flight data to see if each avionics component is behaving according to the specification. 
This process of testing requires control of the UAV, continuously deciding various actions, and observing state change in the UAV and environment.  
To automate this process, human-level decision-making and control of the UAV are required at runtime. 
Such a problem can be devised in the form of a Markov Decision Process (MDP)~\cite{aibook}.
In this case, an AI technique, deep reinforcement learning (DRL) is an appropriate choice compared to meta-heuristics search algorithms. 
Moreover, DRL provides a mechanism for on-the-fly testing of UAS in a dynamic environment which cannot be achieved by traditional model-based testing approaches (e.g.,~\cite{b5}).
Using DRL, our approach introduces an \aitester{} in place of a human tester to perform the system-level testing of the UAV.
We formulate this problem as an MDP which consists of a set of states, a set of possible actions that can be performed in a state, a transition model, and a reward function~\cite{aibook}. 
For this purpose, the proposed profiles, UAS structural profile, and UAV behavioral profile play an important role. 
More discussion on the adaptation of DRL to automate the system-level testing of the UAV is given in the following subsections.

\begin{figure*}[!t]
	\centerline{\includegraphics[width=\linewidth,height=\textheight,keepaspectratio]{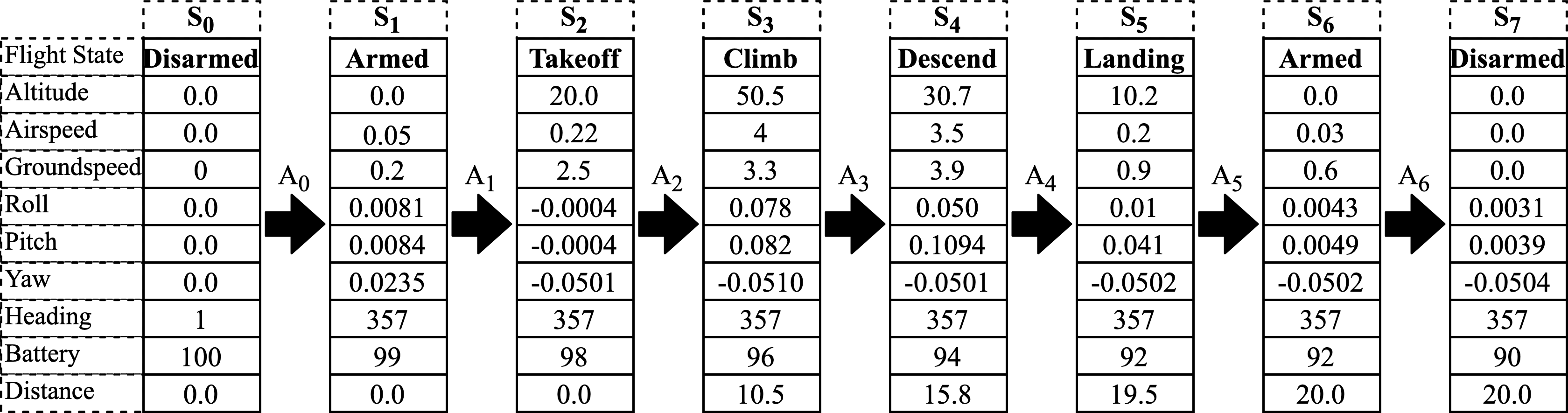}}
	\caption{An example of the UAV flight scenario with different state changes}
	\label{fig:uav-states}
\end{figure*}

\subsubsection{Definitions}\label{definitions}
\noindent\textbf{Definition 1 (State Space).} A state space is defined as an ordered set of states \mbox{$S = \{S_0, S_1, S_2, \dots, S_g\}$} in which each state is represented by a tuple \mbox{$S_x = \tup{s_0, s_1, \dots, s_9}$}. 
Where $S_g$ represents the goal (final) state and $s_0$ represents one of the flight state stereotypes modeled in the UAV flight state machine. 
The remaining state elements are denoted as $s_1 := altitude$, $s_2 := airspeed$, $s_3 := groundspeed$, $s_4 := roll$, $s_5 := pitch$, $s_6 := yaw$, $s_7 := heading$, $s_8 := battery$, and $s_9 := distance$. 
\textcolor{black}{
Given the partially observable environment of UAVs~\cite{xue2023multi}, we consider the state space as a subset of the observation space. 
Hence, our definition of the state space incorporates a selection of properties from the UAS profile (\Cref{fig:uav-profile}), which constitutes the observation space (as per \textit{Definition 5}). 
This state space design is based on key properties identified through various experimental trials. 
We keep the state space size small to avoid the ``curse of dimensionality"---a phenomenon where the amount of data required to learn grows exponentially with the state space size, potentially hindering the agent's learning capability~\cite{bengio2013representation}.  
The main objective is to streamline learning and decision-making processes within high-dimensional environments like UAVs~\cite{sevcik2008testing}. 
}

\Cref{fig:uav-states} shows an example of the different states in the state space considering the UAV flight state machine shown in \Cref{fig:uav-fsm}. 
For the given example, the initial state is represented as \mbox{$S_0 = \tup{\mbox{\em Disarmed}, 0.0, 0.0, 0, 0, 0.0, 0.0, 0.0, 1, 100, 0.0}$}, where $\mbox{\em Disarmed}$ corresponds to $s_0$ in the tuple $S_x$. 
In this state, all the values are zero except the battery level which is 100. 
The next state is \mbox{$S_1 = \tup{\mbox{\em Armed}, 0.0, 0.05, 0.2, 0.0081, 0.0084, 0.0235, 357, 99, 0.0}$} where the \emph{airspeed}, \emph{groundspeed}, \emph{roll}, \emph{pitch}, \emph{yaw}, \emph{heading}, and \emph{battery} values are little changed. 
After the UAV is armed, the next flight state is $\mbox{\em Takeoff}$. 
Therefore, the next state is \mbox{$S_2 = \tup{\mbox{\em Takeoff}, 20.0, 0.22, 2.5, -0.004, -0.004, -0.0501, 357, 98, 0.0}$} in which the \emph{altitude} is increased to 20.0m, the \emph{airspeed} is increased to 0.22, the \emph{groundspeed} is increased to 2.5, and there is a small change in \emph{roll}, \emph{pitch}, \emph{yaw}, and \emph{battery} values. 
Similarly, the UAV state continues to change i.e., from $S_2$ to $S_3$ to $S_4$ to $S_5$ to $S_6$ to $S_7$ state which is the goal (final) state.
The goal state $S_g$ is represented as \mbox{$S_7 = \tup{\mbox{\em Disarmed}, 0.0, 0.0, 0.0, 0.0031, 0.00039, 0.0504, 357, 90, 20.0}$}.

\vspace{5pt}\noindent\textbf{Definition 2 (Action Space).} The action space is represented as a set $A = \{A_0, A_1, A_2, \dots, A_n\}$, where \emph{n} is the total number of events in the UAV flight behavior model. 
Each action in the action space $A$ corresponds to an event modeled in the flight behavior model of the UAV corresponding to abstract flight events of the UAV behavioral profile.  
For the example state transitions shown in \Cref{fig:uav-states}, the action $A_0$ corresponds to the event \emph{armUAV()}, the action $A_1$ corresponds to the event \emph{takeoff()}, the action $A_2$ corresponds to the event \emph{increaseAlt()}, the action $A_3$ corresponds to the event \emph{decreaseAlt()}, the action $A_4$ corresponds to the event \emph{landUAV()}, the action $A_5$ corresponds to the event \emph{armUAV()}, and the action $A_6$ corresponds to the event \emph{disarmUAV()}.

\vspace{5pt}\noindent\textbf{Definition 3 (Correct/Incorrect Action).} An action is termed as correct if it is one of the possible actions (modeled in the flight state machine) for a given current state, otherwise, it is termed as an incorrect action. 
Our approach uses the UAV flight state machine to distinguish between correct and incorrect actions. 
The UAV flight state machine provides the possible set of corrective actions in a particular flight state.
The distinction between correct and incorrect action is important in our case because if the selected action is incorrect it will either create zero or a negative effect on the UAV. 
For example, if the UAV is on the ground in the \emph{\mbox{Armed}} state and the selected action is \emph{Loiter()}, it is not possible for the UAV to start loitering.

\vspace{5pt}\noindent\textbf{Definition 4 (Transition Model).} A transition model is defined as a probability of transition from a current flight state to the next flight state (in states space $S$) using an action from the actions space $A$. 
\Cref{eq:tm} shows a transition model in which $S_{x+1}$ represents the next state in $S$, 
$S_x$ represents the current state in $S$, and $A_x$ represents the current action in $A$. 
According to the equation, there is a probability of the transition from the current state $S_x$ to the next state $S_{x+1}$ using the action $A_x$. 

\begin{equation}\label{eq:tm}
	T = P(S_{x+1} | S_x, A_x)
\end{equation}

For the example state transitions diagram shown in \Cref{fig:uav-states}, using the action $A_0$ in state $S_0$, the next state is $S_1$. 
Similarly, the action $A_1$ in state $S_1$ leads to the next state $S_3$.  

\vspace{3pt}\noindent\textbf{Definition 5 (Observation Space).} An observation space is defined as a vector of UAV properties, i.e., $O = [o_0, o_1, o_2, \dots, o_n]$ from the UAS SUT domain model, created using the UAS structural profile. 
Where $o_x$ corresponds to the UAV properties such as \emph{thrust}, \emph{airspeed}, \emph{vx}, \emph{vy}, and \emph{vz}.

\subsubsection{Reward Function}\label{sbsec:reward}
The objective of \aitester{} is to take control of the UAV flight during simulation, perform various actions on the UAV, and maximize the number of violations of the expected behavior. 
The reward function is designed based on two parameters, (i) correct/incorrect actions, and (ii) violations of the expected behavior specified as OCL constraints. 
These two parameters are important because a series of correct/incorrect actions may lead a UAV to crash. 
For example, if \aitester{} continues to execute a correct action \emph{decreaseAlt()} multiple times in \emph{\mbox{Descent}} state, the UAV can hit the ground leading it to crash. 
This can happen frequently during training because the inherent behavior of the DRL algorithm involves exploring various possibilities of actions and observing their outcomes. 
Therefore, a correct action leading the UAV to crash is not considered a fault in the UAV. 
To handle such a situation, our approach utilizes the deviations from a UAV's expected behavior based on violations of OCL constraints. 
The trace of OCL violations (i.e., deviations from the expected behavior) can support diagnosing faults in the SUT.  
Since OCL constraints are specified corresponding to each flight state of the UAV, it is possible to have multiple violations of OCL constraints at different states during the whole flight. 

The reward for a single action performed in a particular state is calculated using \Cref{eq:reward}. 
For a correct action, the \aitester{} gets a reward of $(1+m)$, where $1$ represents the reward for the correct selection of action and $m$ represents the number of violations in the expected behavior (failed OCL constraints) based on that action. 
To calculate the number of failed OCL constraints, the UAV flight data (e.g., altitude, airspeed, roll, pitch, and yaw) sent by the state observer is used to populate an instance model which is generated using the input UAS SUT domain model (as shown in \Cref{fig:ate-app}).  
The instance model populated in this way represents the current flight context of the UAV during execution. 
The constraints evaluator uses the input OCL constraints and state invariants for the current flight state and evaluates these constraints on the UAV instance model. 
The constraints evaluator returns the number of failed OCL constraints (\emph{m}) for the given flight data. 
Since different OCL constraints are specified on a particular state, multiple violations of one constraint are considered as one violation. 
In this way, the \aitester{} learns the flight behavior of the UAV, and at the same time, utilizes the flight behavior to explore the scenarios that violate the expected behavior of the SUT. 
For an incorrect action, the \aitester{} gets the reward of $-1$.

\begin{equation}\label{eq:reward}
  r = \left\{ 
  \begin{array}{ l l }
    1+m & \quad correct\ \ action \\
    -1 & \quad incorrect\ \ action
  \end{array}
\right.
\end{equation}

\begin{equation}\label{eq:reward2}
	R = \sum_{t=0}^{n} \gamma^{t} * r_t
\end{equation}

The cumulative reward for a single episode is calculated using \Cref{eq:reward2}. 
According to this Equation, the cumulative reward is the added sum of the reward for individual action performed at time \emph{t} (calculated using \Cref{eq:reward}) multiplied with discount factor ${\gamma}^t$ at time step \emph{t}. 
The discount factor $\gamma$ is between 0 and 1, i.e., $\gamma \in [0, 1]$.
\textcolor{black}{
This factor is essential in determining the extent to which the \aitester{} prefers future rewards over immediate ones. 
If the discount factor is closer to 1, the \aitester{} gives more preference to future rewards. 
This means the \aitester{} will explore more of the state space, even if immediate rewards are not apparent. 
This can lead to broader state space coverage as the \aitester{} is incentivized to explore and comprehend as much of the environment as possible. 
On the other hand, if the discount factor is closer to 0, the \aitester{} is more inclined towards immediate rewards. 
This can limit the exploration of the state space. 
As a result, the \aitester{} may not thoroughly explore the environment, possibly missing out on regions of the state space that could provide higher long-term rewards. 
Therefore, to enable broader state space coverage and thorough environment exploration, the discount factor should ideally be adjusted closer to 1, aligning with similar practices in the domain~\cite{mnih2015human}. 
This value will progressively decrease after each time step \emph{t}, thereby systematically shifting \aitester{}'s focus toward immediate rewards.
Consequently, the \aitester{} initially prefers future rewards and gradually transitions its focus towards immediate rewards as time progresses.
}

\subsubsection{\textcolor{black}{Determining LSTM Architecture for \aitester{}}}
The UAV flight data at a particular time is correlated to the flight data at the previous instance of time. 
Due to this, UAV flight data holds a temporal relationship based on environmental context which is highly dynamic. 
Retaining information among these data points in subsequent time steps is important to comprehend the flight context. 
Moreover, our approach utilizes Deep Q-Network (DQN)~\cite{mnih2015human} which works on the experience replay method by taking random samples from the replay memory of past experiences. 
This is an inherent feature of DQN to make learning efficient, however, it breaks correlation among different data points which makes it difficult to acquire the flight context. 
In such a scenario, an LSTM network is a suitable choice due to its ability to handle sequential data and long-term dependencies. 

\begin{figure}[ht]
    \centerline{\includegraphics[width=12.6cm,height=4.6cm,keepaspectratio]{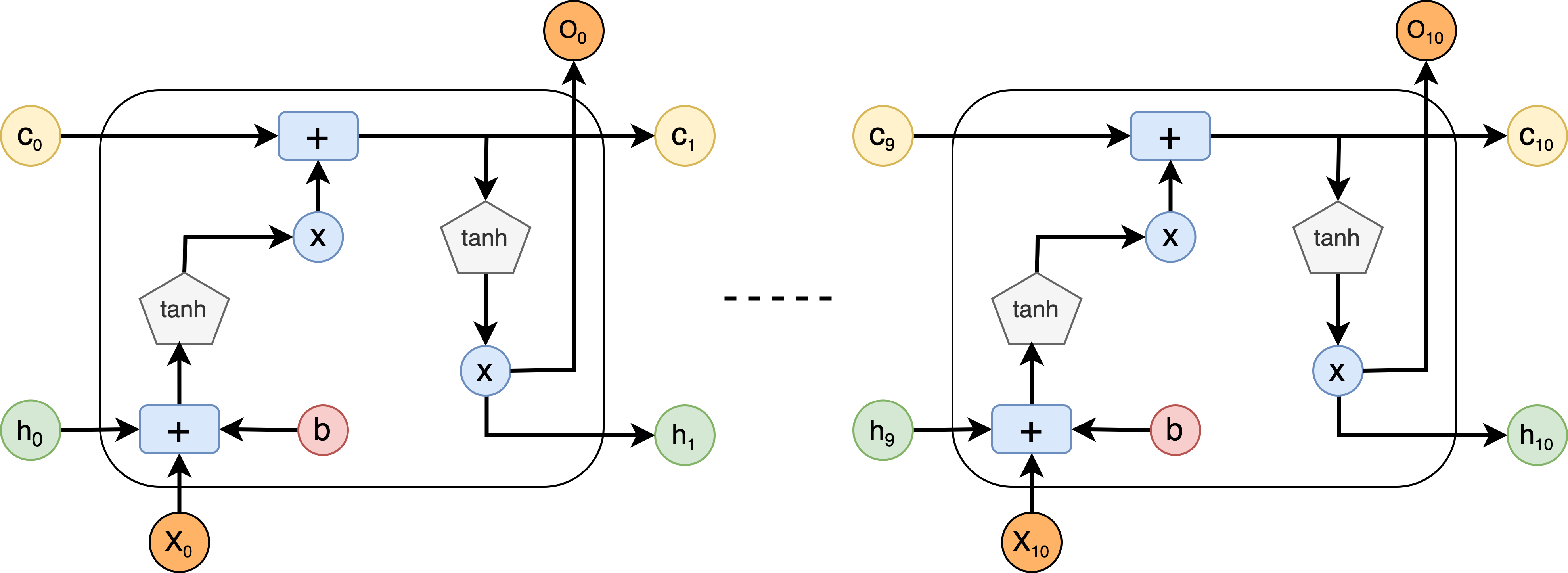}}
	\caption{A single layer representation of LSTM architecture for \aitester{}}
	\label{fig:ann}
\end{figure}

\Cref{fig:ann} shows a single-layer representation of LSTM architecture used in our approach. 
The LSTM network architecture used in our approach consists of an input layer, three LSTM layers with a hidden dimension of 10, and an output layer. 
The inputs of the neural network represent a state in states space $S$. 
The first input is the abstract flight state that corresponds to one of the states in the flight behavior model. 
The remaining inputs such as \emph{altitude}, \emph{airspeed}, \emph{groundspeed}, \emph{roll}, \emph{pitch}, \emph{yaw}, \emph{battery}, \emph{heading}, and \emph{distance} represent the information displayed on the cockpit of the ground control station which is received from the UAV during flight. 
This information is necessary for the remote pilot to control the UAV during flight~\cite{sartajtesting}. 
That is, based on this information, the remote pilot decides a sequence of actions to successfully fly the UAV. 
The activation function used for the hidden layers of our LSTM network is the default hyperbolic tangent (tanh)~\cite{maas2013rectifier}. 
The output of the neural network is the possible actions (from the action space $A$) that can be performed given the input state. 
From the possible actions, the action with the highest probability is selected using the \emph{softmax} function.

\subsubsection{UAV Environment}
The agent has to interact with the UAV to perform an action, observe the state of the UAV, and get the reward for the selected action. 
To facilitate these operations, we build a goal-based environment with the goal of successfully landing the UAV on the ground. 
Under normal conditions, after performing different flight operations, if the UAV should land on the ground, the goal is achieved. 
Another possibility is that the UAV crashes due to a correct/incorrect sequence of actions (\Cref{sbsec:reward}). 
In such a case, the goal is not achieved.
UAV crashes during test execution can be observed in a simulation environment without requiring an actual UAV.

The environment takes state space and action space from the flight behavioral model (\Cref{sbsec:fbm}) and the observation properties from the domain model (\Cref{sbsec:dm}). 
For each action in the action space, the environment implements the UAV-specific action commands. 
Similarly, for each action in a particular state, the environment keeps track of the next state. 
If the next state is the final (goal) state, the environment notifies the agent that the goal has been achieved. 
Based on the action (correct/incorrect), the environment produces observations for the next state. 
During the execution, it receives the action from the agent, determines the current state, performs the action on the UAV in simulation, observes the next state, and checks if the goal state is achieved or not.

\subsubsection{Testing Process of \aitester{}}\label{sec:aitester}
To begin, \aitester{} requires three inputs from avionics testers (as shown in \Cref{fig:ate-app}); (i) the domain model of the UAS SUT developed using the UAS structural profile, (ii) the flight behavior in the form of a UML state machine developed using the UAV behavioral profile, and (iii) the expected behavior of UAS SUT in the form of OCL constraints. 
The UAS SUT domain model is used to initialize the observation space and to generate an instance model that supports the evaluation of OCL constraints during execution. 
The flight behavior model of the UAV is used to initialize the state and action space of the DRL algorithm.

\aitester{} incorporates Deep Q-Network (DQN) as a DRL algorithm~\cite{mnih2015human}. 
DQN has been widely used to train different game-playing bots and to get human-level control~\cite{mnih2013playing, jaderberg2019human}. 
The testing process starts by initializing the LSTM network called the policy network, the environment, and by getting the initial state of the UAV. 
For the selected state, the \aitester{} selects an action randomly (explore) or by using the policy network (exploit). 
For this purpose, the \aitester{} uses Epsilon $\epsilon$-greedy strategy to balance exploration and exploitation. 
Before executing the selected action, it is analyzed whether this action can be performed on a UAV or not.
This is done using the flight state machine. 
If the selected action can not be performed, it is omitted and the negative reward is returned. 
In this case, the state of the UAV is not changed. 
If the selected action is fine, it is performed on a UAV operating in an environment.

Using the UAV flight states and actions, the \aitester{} starts exploring various flight states using different actions (e.g., \emph{\mbox{takeoff}} or \emph{\mbox{land}}). 
The \aitester{} selects an action and executes it using \textit{Action Executor}. 
The \textit{Action Executor} receives an action, establishes a connection with the UAV through DroneKit API, and sends the action command to the UAV. 
The function of DroneKit is to receive the action command for the UAV, send the command to the UAV using a communication link (using Mavlink), and receive the state of the UAV. 
The continuous change in the flight state of the UAV is monitored by a \textit{State Observer}. 
This helps the \aitester{} to keep track of the current state, action, and the next state of the UAV. 
Moreover, the flight data observed during execution is used to populate an instance model of the UAS SUT domain model. 
The instance model is evaluated on the OCL constraints and state invariants using a \textit{Constraints Evaluator}. 
The \textit{Constraints Evaluator} provides the number of violations of OCL constraints. 
Based on the selected action and the number of violations of OCL constraints, the \aitester{} gets the reward.

During training, this process continues for a specified number of episodes, and the \aitester{} continues to explore different flight states based on different actions and learns to violate OCL constraints (expected behavior of the SUT). 
The training is performed using a flight simulator. 
At the end of the training process, the trained model is saved. 
In the evaluation phase, the trained model is loaded and \aitester{} selects various actions using the trained model, performs the selected actions on the UAV operating in an environment, observes the outcomes, and compiles testing results for further analysis by avionics testers. 
The compiled results include information related to UAV flight states, actions, and the number of failed OCL constraints for each execution. 
This information helps an avionics tester to analyze the sources of faults in the SUT. 
At the end of the process, the policy network is saved so that it can be reused in later stages of testing.

\section{Implementation}\label{sec:tool}
We implemented the proposed approach in a toolset to support the automated system-level testing of the UAV. 
All main modules of the toolset are packed into two separate packages. 
The toolset is available at the GitHub repository\footref{uast-repo}. 
In the following subsections, we discuss every module of the toolset w.r.t. each package.

\subsection{Model Handler}
This component takes input UAS SUT domain model, flight behavior model, and OCL constraints. 
Using this input, it loads the models, generates a model instance, prepares the instance model for evaluation during execution, and evaluates the OCL constraints on the instance model. 
Each module of this component is developed using the Java programming language. 

\subsubsection{Model Loader}
First, this module takes the UAS SUT domain model in the form of UML and loads this model in a data structure along with the meta-class elements (i.e., profile stereotypes). 
Second, it takes the flight behavior model in the form of a UML state machine and traverses the model to extract flight states, events, and transitions. 
In case the state machine has composite and orthogonal states, these states are flattened using the algorithm presented by Binder~\cite{b5}.
For the modeling manipulation, it uses an external API named Eclipse Modeling Framework (EMF\footnote{https://www.eclipse.org/modeling/emf/}) that allows it to load and read the model information conveniently.  
The third function of this module is to load the OCL constraints and state invariants.

\subsubsection{Instance Generator}
This module uses the domain model provided by \textit{Model Loader} and the UAS profile to generate an instance model. 
For this purpose, EMF is used to create instances of the domain classes and the associations among these classes. 
The properties for each class are created using the domain class and profile stereotype attributes. 
A slot is created for each property in the instance model.  
The generated instance model is saved in the form of a UML model. 

\subsubsection{Instance Populator}
The responsibility of this module is to receive the UAV flight data and populate the instance model generated using \textit{Instance Generator}. 
First, the classes and properties are mapped between the UAV flight data and the instance model. 
After that, the flight data corresponding to each UAV class and properties are used to fill the corresponding slots in the instance model.

\subsubsection{Constraints Evaluator}
The core functionality of this module is to evaluate the OCL constraints and state invariants on the instance model containing the UAV flight information. 
For this purpose, the Eclipse OCL\footnote{https://projects.eclipse.org/projects/modeling.mdt.ocl} evaluator is used. 
Before starting the evaluation process, the OCL constraints and state invariants are extracted (using \textit{Model Loader}) corresponding to each state of the UAV flight. 
During execution, the flight data received from the environment is sent to \textit{Instance Populator}. 
The instance model is used to prepare an OCL evaluator environment for the evaluation of OCL constraints~\cite{sartaj2019search,sartaj2024efficient}. 
Before starting the evaluation process, OCL constraints are loaded from a single file corresponding to each state of the UAV flight. This file contains general constraints and state-specific constraints, resembling the format depicted in \Cref{lst:constraints}. 
All OCL constraints and state invariants are evaluated on the populated instance model. 
If an OCL constraint or state invariant fails on the instance model, the Eclipse OCL evaluator returns \emph{false}. 
If the OCL constraint or state invariant is passed on the instance model, the Eclipse OCL evaluator returns \emph{true}. 
At the end of the constraints evaluation, the evaluator returns the total number of failed OCL constraints (violations in the expected behavior of the SUT) for the given flight data.

\subsection{\aitester{}}
This constitutes the primary component of the toolset. 
The key functions of this component are UAV environment preparation, interaction with \textit{Constraints Evaluator} for the reward calculation, and training and evaluation modes for system-level testing of UAVs. 
Since \aitester{} is aimed to be used at different levels of testing, the tool contains a training module as well as an evaluation module. 
All modules of the \aitester{} are developed using Python programming language. 
The deep learning framework used for this purpose is PyTorch~\cite{paszke2017automatic}.

\subsubsection{Training Module}
The main function of this module is to train the \aitester{} for the automated system-level testing of the UAV. 
This module implements a deep reinforcement learning algorithm named Deep Q-Network (DQN) along with the LSTM network. 
For the DQN, it takes various hyperparameters such as the number of episodes, batch size, gamma, replay memory size, target update, $\epsilon$-start, $\epsilon$-end, and $\epsilon$-decay. 
Before the start of the training process, it determines the current state of training i.e., whether the training needs to start from scratch or load the previous checkpoint and start from that point. 
To reload from the previous checkpoint, the LSTM networks (both policy and target network) are loaded using PyTorch model loading API. 
Each episode of the training process starts with an initial state, selects an action using the LSTM network, gets a reward from the \textit{Environment}, and observes the next state. 
During each episode, the explored states, correct/incorrect actions, and resultant failed OCL constraints are compiled and stored in a file. 
To balance exploration and exploitation, the Epsilon $\epsilon$-greedy strategy is implemented. 
For the training, Adam optimizer~\cite{kingma2014adam} and SmoothL1Loss functions are used. 
After a specified interval, the state of the training process (including LSTM networks) is stored. 
When the training process finishes, the trained LSTM network is stored in the specified directory so that it can be used for evaluation.

\subsubsection{Evaluation Module}
The role of this module is to perform the system-level testing of UAVs when the \aitester{} is trained in the SIL simulator. 
This module takes a trained LSTM network and loads it using PyTorch. 
The process of this module is identical to the main procedure of the DQN algorithm during training except for the learning parameters such as the use of replay memory, calculation of Q-values, loss calculation, and gradient optimization. 
The main use of this module is the facilitation of testing at HIL and PIL levels. 
This module also supports the compilation of testing results at the end of each test execution.

\subsubsection{Environment}
The core feature of this module is to create an environment in which the \aitester{} can perform actions, observe the UAV flight state, and calculate rewards. 
For this purpose, this module provides an environment manager to facilitate these operations. 
To perform an action on the UAV, this module uses the DroneKit API\footnote{https://dronekit.io/}. 
The DroneKit uses Mavlink as a communication link to communicate with the UAV. 
The main benefit of the DroneKit is that it is independent of a particular simulator environment. 
This makes our toolset capable of performing system-level testing of the UAVs at SIL, HIL, and PIL levels.

\subsubsection{Py2Java Communicator}
In this module, py4j\footnote{https://www.py4j.org} API is used to create a communication channel between the python component (\textit{\aitester{}}) and the Java component (\textit{Model Handler}). 
The core feature is to send the UAV flight information to \textit{Model Handler} for the constraints evaluation, receive the results, and pass them on to the \textit{\aitester{}} during the execution. 
Initially, this module gets the UAS SUT domain model details and flight behavior model information from the \textit{Model Handler}. 
This information is used to create and initialize the state, action, and observation space of the environment.

\subsubsection{UAV Commands API}
We provide an API for the execution of UAV flight events on the simulator environment. 
This API contains the concrete implementation of DroneKit-specific commands for ArduCoper and ArduPlane corresponding to the abstract flight events modeled in the flight behavior profile. 
Since the API contains the implementation of abstract flight events, it can easily be extended and modified for any other simulator that supports DroneKit.

\section{Empirical Evaluation}\label{sec:eval}
This section presents the empirical evaluation of \aitester{} including two experiments. 
The first experiment was performed using the ArduCopter autopilot system and the second experiment was conducted using an industrial case study of ground control station (GCS) cockpit display systems (CDS), i.e., GCS-CDS. 
In the following subsections, we discuss research questions, case studies, both experiments' setups, the design and execution of the experiments, the results of both experiments, and the possible threats to the validity of experimental results. 

\subsection{Research Questions}
\aitester{} utilizes a UAV flight state machine for the system-level testing of the UAS system under test (SUT) with the goal of violating the expected behavior of the SUT. 
There are two key aspects for analyzing the performance of the \aitester{}.  
The first is the effectiveness of \aitester{} in generating test scenarios that can lead to deviations from the expected behavior of the UAS SUT. 
During the testing process, the second aspect is the ability of the \aitester{} to explore diverse paths from the UAV flight state machine. 
We formulate the following research questions for the experimental evaluation based on these two aspects. 

\begin{itemize}[leftmargin=10pt]
	\item \textbf{RQ1:} Is \aitester{} effective in violating the expected behavior of a UAS SUT?
	\item \textbf{RQ2:} Is \aitester{} capable of exploring diverse flight paths during testing?
\end{itemize}

Both RQs are important from the testing perspective of a UAS SUT. 
The goal of the \aitester{} is to perform system-level testing of a UAS SUT using a state machine representing the flight behavior of a UAV. 
RQ1 assesses \aitester{} in terms of violations of the expected behavior of a UAS SUT that can lead to potential faults. 
This is a primary concern of testing a UAS.
Since a flight state machine is utilized in the testing process, it is quite possible that \aitester{} learns to explore a specific set of flight states and actions leading to maximum violations using the same set of OCL constraints. 
This results in testing the same behavior multiple times which is not a desired feature.
Therefore, the analysis of diversity in exploring flight paths is an important consideration.

\subsection{Case Studies Description}
For the first experiment, we used ArduCopter\footnote{https://ardupilot.org/copter/} which is a part of the widely used open-source ArduPilot\footnote{https://ardupilot.org/ardupilot/index.html} project. 
The ArduCopter supports operating a variety of multi-copter UAVs including helicopter, tri-copter, quad-copter, hexacopter, etc. 
It supports many flight modes including manual, semi-autonomous, and fully autonomous. 
Its software is compatible with a wide range of hardware e.g., Pixhawk, CUAV v5 Plus, OpenPilot Revolution, etc. 
It has also been used for research purposes in various experimental settings~\cite{ma2019testing,koch2019reinforcement}.

\begin{figure*}
	\centering
	\subfigure[Airspeed Indicator (ASI)]{\includegraphics[width=3.9cm,height=3.9cm]{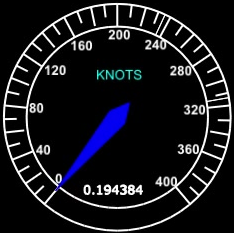}\label{first}}
	\subfigure[Altimeter]{\includegraphics[width=3.9cm,height=3.9cm]{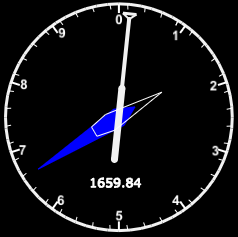}\label{second}}
	\subfigure[Turn Coordinator]{\includegraphics[width=3.9cm,height=3.9cm]{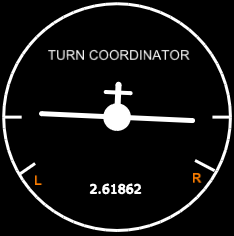}\label{third}}
	\caption{Three important flight instruments of GCS-CDS}\label{gcscds}
\end{figure*}

For the second experiment, we used an industrial case study representing the CDS of GCS (i.e., GCS-CDS). 
The GCS-CDS contains three important flight instruments as shown in \Cref{gcscds}. 
The first flight instrument represents an airspeed indicator (ASI). 
\Cref{first} shows an ASI displaying the airspeed value of $\approx$0 knots. 
The second flight instrument represents an altimeter.  
\Cref{second} shows an altimeter indicating the altitude value of $\approx$1660 feet above sea level (ASL). 
The third flight instrument represents a turn coordinator that is used to display the current turning direction and the bank angle of the UAV. 
\Cref{third} shows a turn coordinator displaying the turning direction towards the right and the bank angle of $\approx$3 degrees.

\begin{table}[!t]
	\centering
	\caption{Modeling statistics of both case studies}
	\begin{tabular}{|l|l|l|}
		\hline
		\textbf{Modeling Artifact} &\textbf{ArduCopter}&\textbf{GCS-CDS}\\
		\hline
		UAV model classes &6&41\\\hline
		UAV model properties & 23&216\\\hline
		State machine states &12&14\\\hline
		State machine events & 23&25\\\hline
		State machine transitions &43&45\\\hline
		UAS profile applied stereotypes &66&40\\\hline
		OCL constraints&48&30 \\\hline
	\end{tabular}
	\label{tab:modelstats}
\end{table}

\subsection{Experiments Setup}
The modeling statistics of both case studies are shown in \Cref{tab:modelstats}. 
For the first experiment using ArduCopter, we used the proposed UAS profile to model the domain and behavior of the ArduCopter. 
The ArduCopter domain model consists of six domain classes and 23 properties. 
The ArduCopter behavioral model comprises 12 flight states, 23 events, and 43 transitions. 
The domain model is shown in \Cref{fig:uav-dm} and the flight behavioral model is shown in \Cref{fig:uav-fsm} and \Cref{fig:uav-tsm}.
The total number of applied stereotypes is 66. 
The expected behavior of the quadcopter is specified in the form of OCL constraints that are 48 in number (including general and state invariants). 
We used the ArduPilot software-in-the-loop simulator which was built using the binaries of the same version of the ArduCopter source code. 

For the second experiment using GCS-CDS, we used the UAS profile and CDS testing profile proposed in previous work~\cite{sartajtesting}. 
The UAS profile was used to model the flight behavior of a UAV and the CDS testing profile was used to model the domain of the GCS-CDS. 
The GCS-CDS domain model consists of 41 classes and 216 properties. 
The ArduCopter behavioral model comprises 14 flight states, 25 events, and 45 transitions. 
The total number of applied stereotypes is 40. 
The expected behavior of GCS-CDS is specified in the form of OCL constraints that are 30 in number.

\begin{lstlisting}[label=lst:expconstraints, language=ocl, caption={An excerpt of the OCL constraints used in our experiments}, linewidth=13cm, numbers=none]
-- minimum battery level during Armed state
C1: context ArduCopter inv: self.oclIsInState(Armed) and self.battery.level>10
-- copter's roll values range during PositionHold state 
C2: context ArduCopter inv: self.oclIsInState(PositionHold) and self.attitude.roll>=0 and self.attitude.roll<=5
-- copter's yaw and yaw_rate ranges during Circle state 
C3: context ArduCopter inv: self.oclIsInState(Circle) and self.attitude.yaw>=-30 and self.attitude.yaw<=30
C4:context ArduCopter inv: self.oclIsInState(Circle) and self.attitude.yaw_rate>=0 and self.attitude.yaw_rate<=5
-- airspeed values range during Approach state
C5: context ArduCopter inv: self.oclIsInState(Approach) and self.airspeed>0 and self.airspeed<50
\end{lstlisting}

\textcolor{black}{
The constraints for our experiments were identified through a careful analysis of the Ardupilot documentation, as well as consultations with avionics testers to consider important testing scenarios.
\Cref{lst:expconstraints} presents a selection of OCL constraints utilized in our experiments. 
Constraint \textit{C1} defines the minimum battery level for ArduCopter in \mbox{\emph{Armed}} state, aiming to ensure the UAV has sufficient battery to take off. 
If ArduCopter initiates takeoff with a low battery level and further depletes its battery during flight without sufficient power for a failsafe landing action, it could potentially result in a crash. 
Constraint \textit{C2} specifies roll angle range during \mbox{\emph{PositionHold}} state to maintain the stability of the ArduCopter. 
If the roll angle surpasses the defined limit, it could destabilize the ArduCopter, potentially leading to collisions with obstacles. 
Constraints \textit{C3} and \textit{C4} define ranges for yaw angle and yaw rate during \mbox{\emph{Circle}} state. 
Both yaw angle and rate are crucial in this state because violating these limits could potentially result in uncontrollable rotations, flipping over, loss of control, and possible collisions with obstacles. 
Constraint \textit{C5} specifies airspeed values range during \mbox{\emph{Approach}} state. 
Exceeding the airspeed limit during this state may lead to a rapid and uncontrolled landing, or hit the ground with extreme force, potentially causing damage to both the copter and its surroundings. 
It is important to note that in our approach, these constraints do not force a UAV into unsafe scenarios. 
They mainly define the expected ranges for a specific flight state, and exceeding these ranges could signify deviations from the expected behaviors. 
}

\begin{table}[!t]
	\centering
	\caption{Hyperparameter values used for both experiments}
	\begin{tabular}{|l|l|}
		\hline
		\textbf{Hyperparameter} &\textbf{Value}\\
		\hline
		Learning rate &0.001\\\hline
		Batch size &128\\\hline
		Replay memory size &1024\\\hline
		Target update &10\\\hline
		Gamma ($\gamma$) & 0.999\\\hline
		$\epsilon$-start &1\\\hline
		$\epsilon$-end &0.01\\\hline
		$\epsilon$-decay &100\\\hline
        Training episodes &1000\\\hline
        Evaluation episodes &100\\\hline
	\end{tabular}
	\label{tab:hyperparams}
\end{table}

\subsection{Experiments Execution}

For the training phase, we used hyperparameters presented in \Cref{tab:hyperparams}. 
Since there are no standard guidelines for selecting hyperparameters for training, we selected these hyperparameters based on the various experimental trials with different values of parameters and their combinations. 
We executed \aitester{} and Random to analyze the training progress considering the randomness factor.
Both \aitester{} and Random approaches were executed on two machines with the same specifications, i.e., core i7 3.6 GHz processor, 64 GB RAM, 1 TB hard drive, NVIDIA GeForce\textregistered RTX 2080 Ti graphics card, and Windows 10 64-bit operating system. 
We set the number of episodes to 1000 for \aitester{} and the same number of iterations for Random. 
At the end of the training session, a trained model for \aitester{} was stored.

For the first experiment with ArduCopter, we used the trained \aitester{} and ran it 100 times (\Cref{tab:hyperparams}). 
We also ran Random 100 times for the comparison. 
We used random as a baseline because the available approaches in literature either require manual test scenario creation, focus on testing a specific subsystem of UAV autopilot, or do not provide public tool support. 
Moreover, avionics testers commonly execute UAV flight scenarios with random actions and data using a flight simulator, a practice similar to random testing aimed at identifying system faults. 
Alternatively, if the testing focus is on uncovering security vulnerabilities, fuzz testing could serve as a comparative approach. While fuzz testing is security-focused~\cite{zhu2022fuzzing}, aiming to identify vulnerabilities through invalid or unexpected inputs, our evaluation does not aim for security testing of the UAS.

For the second experiment using GCS-CDS, we compared \aitester{} with the baseline model-based CDS testing approach~\cite{sartajtesting}.
For the CDS testing approach, we repeated the previous experiment procedure~\cite{sartajtesting}. 
This experiment was executed on two machines with the same specifications used in the first experiment. 
We set the number of episodes to 100 (\Cref{tab:hyperparams}) for \aitester{} and the same number of paths are used for the CDS testing approach. 
The overall execution time for each case study, across 100 episodes, was approximately 300 minutes. 
It is important to note that the execution time varies based on the machine specifications and the flight simulator being used. 
While we experimented with ArduPilot -- a widely used open-source simulator -- \aitester{} is not tied to any specific simulator. 
It is adaptable and can seamlessly integrate with a faster simulator.

To make our experimental results reproducible, we provide the toolset with the experiment setup and execution information at an online repository\footref{uast-repo}.

\subsection{Experiments Results and Analysis}
To analyze the training progress, we calculated the moving average reward for all episodes. 
The moving average reward (MAR) for \emph{N} episodes ($MAR_N$) is calculated using \Cref{MAR}, where \emph{n} is the number of reward values to consider in one window of the moving average. 
The overall MAR values over total episodes can be calculated using \Cref{MAR2} for each new average window of \emph{n} reward points. 
In this equation, $r_{n+1}$ is the new reward value and the $r_{n-N+1}$ is the oldest value in the previous episode window.

\begin{equation}\label{MAR}
	MAR_N = \frac{1}{N} \sum_{i=N-n+1}^{N} r_i
\end{equation}

\begin{equation}\label{MAR2}
	MAR = MAR_{N,prev} + \frac{1}{N} \left(r_{n+1} - r_{n-N+1}\right)
\end{equation}

\begin{figure}[ht]
	\centerline{\includegraphics[width=11.7cm,height=7.7cm,keepaspectratio]{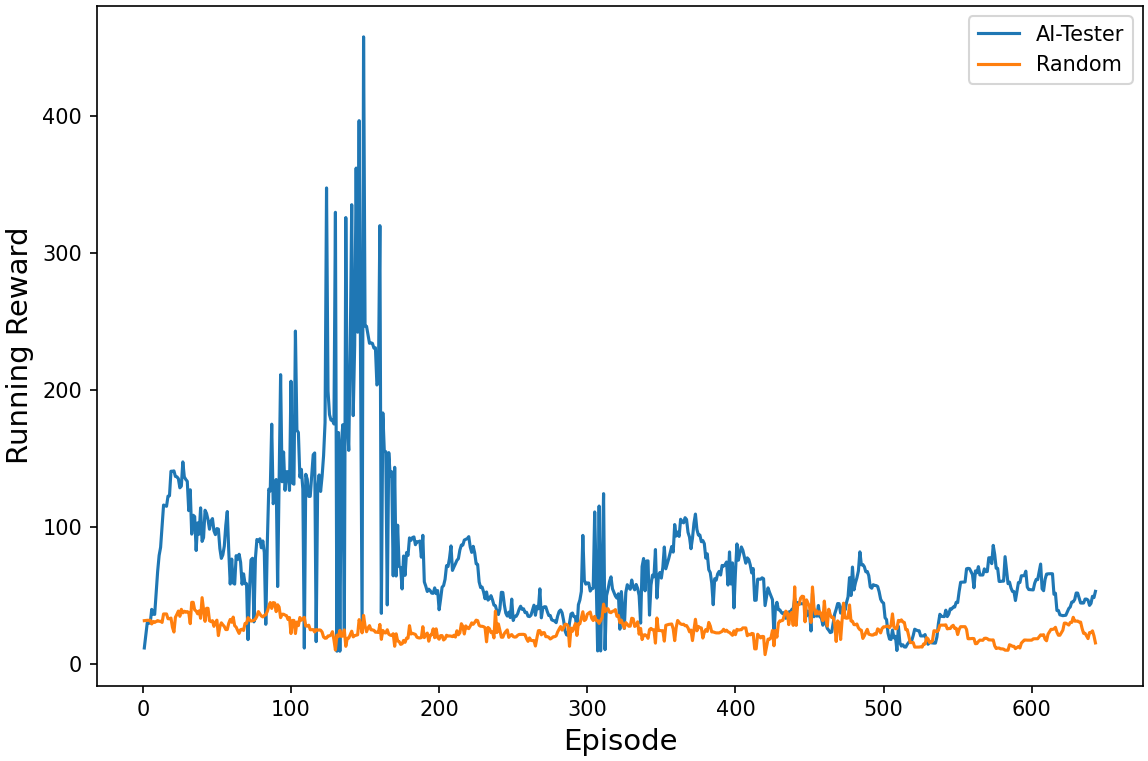}}
	\caption{Analysis of moving average reward for \aitester{} in comparison to Random}
	\label{fig:reward}
\end{figure}

\Cref{fig:reward} presents a comparison between \aitester{} and Random based on the MAR for 1000 training episodes.
The figure demonstrates MAR values (up to $\approx$650) that are computed over 1000 episodes using a larger window size~\cite{ermshaus2023window}, i.e., $n=350$.
It can be observed that the MAR of \aitester{} is high in most of the cases compared to Random. 
MAR of \aitester{} is greater than 50 in a majority of the episodes and in some cases, it is higher than 100, whereas Random maintains MAR $\approx$ 45 in most of the episodes. 
The MAR of \aitester{} and Random is close in some cases.  
This happens when the \aitester{} tries to explore (select random actions) the environment to create a balance between exploration and exploitation. 
From the training outcomes, it can be seen that \aitester{} learns to perform the correct sequence of actions while violating the expected behavior of the SUT. 

To answer RQ1, we performed a comparison based on the deviations from the expected behavior, considering violations of OCL constraints. 
As \aitester{} utilizes UAV flight state machines to explore different paths in a state machine, an important consideration is the exploration of diverse state machine paths during testing. 
To analyze the diverse exploration of UAV flight states, we used path diversity as a metric that was employed in a similar context and different domains, e.g., for Web~\cite{fard2013feedback}. 
For RQ2, we analyzed the statistically significant difference between the two approaches based on path diversity. 
We calculated path diversity using the method defined for state graphs~\cite{fard2013feedback}).

To analyze the results for RQ2 and perform a statistical comparison, we followed the guidelines defined by Arcuri and Briand~\cite{arcuri2011practical}. 
We used the Wilcoxon signed-rank test with a significance level ($\alpha$) set at 0.05. 
The Wilcoxon signed-rank test is a non-parametric statistical test utilized for comparing two different sample sets. 
This test produces a \emph{p-value}, which is used to determine if the difference between two samples holds statistical significance.
A smaller \emph{p-value} (normally $<=\alpha$) suggests that there is a statistically significant difference between the two samples. 
A larger \emph{p-value} (normally $>\alpha$) indicates that there is no statistically significant difference between the two samples. 
To measure the magnitude of the difference, particularly to determine which sample is better, we utilized Cliff's Delta~\cite{cliff1993dominance}.
Cliff's Delta is a non-parametric effect size measure that quantifies the degree of difference between two samples. 
A value of 0 signifies that the two samples are identical. 
A positive value (up to +1) suggests that the first sample is better than the second sample. 
Conversely, a negative value (down to -1) indicates that the second sample is better than the first sample.
The discussion on experiment results corresponding to each research question is given below.

\begin{table*}[!t]
	\centering
	\noindent
	\caption{Comparison results from 100 episodes of the \aitester{} approach and Random Testing baseline for ArduCopter case study}
	\begin{tabular}{l M M M M}\toprule
		\multicolumn{1}{l }{\textbf{}} & \multicolumn{2}{c }{\textbf{\aitester{} Approach}} & \multicolumn{2}{c }{\textbf{Random Testing}} \\ 
		\cmidrule(lr){2-3}
		\cmidrule(ll){4-5}
		\multicolumn{1}{ l }{\textbf{States}} & \textbf{OCL Violations} & \textbf{Unique OCL Violations}& \textbf{OCL Violations} & \textbf{Unique OCL Violations}\\ 
		\cmidrule(lr){1-1}
		\cmidrule(lr){2-3}
		\cmidrule(ll){4-5}
		\multicolumn{1}{ l }{\textbf{AltitudeHold}} & 6964 & 4 & 4646 & 3\\
		\multicolumn{1}{ l }{\textbf{PositionHold}}  & 1780 & 4 & 1328 & 4\\
		\multicolumn{1}{ l }{\textbf{Climb}}  & 3820 & 4 & 2232 & 3\\
		\multicolumn{1}{ l }{\textbf{Descent}}  & 19140 & 4 & 2942 & 1\\
		\multicolumn{1}{ l }{\textbf{Landing}}  & 956 & 4 & 362 & 2\\
		\multicolumn{1}{ l }{\textbf{Loiter}}  & 872 & 4 & 374 & 2\\
		\multicolumn{1}{ l }{\textbf{Takeoff}}  & 696 & 4 & 690 & 4\\
		\cmidrule(ll){1-5}
		\multicolumn{1}{ l }{\textbf{\textit{Total}}} & \textit{34228} & \textbf{\textit{28}} & \textit{12574} & \textbf{\textit{19}}\\ 
		\bottomrule
	\end{tabular}
	\label{ardu-results}
\end{table*}

\begin{table*}[!t]
	\centering
	\noindent
	\caption{Comparison results from 100 episodes for ArduCopter and GCS-CDS based on diversity}
	\begin{tabular}{l l p{3.5cm}}\toprule
		\multicolumn{1}{l }{\textbf{}} & \multicolumn{1}{c }{\textbf{ArduCopter}} & \multicolumn{1}{c }{\textbf{GCS-CDS}} \\ 
		\cmidrule(lr){2-2}
		\cmidrule(ll){3-3}
		\multicolumn{1}{ l }{\textbf{Statistical Tests}} & \textbf{\aitester{} vs Random} & \textbf{\aitester{} vs CDST Approach}\\ 
		\cmidrule(lr){1-1}
		\cmidrule(lr){2-2}
		\cmidrule(ll){3-3}
		\multicolumn{1}{ l }{Wilcoxon signed-rank}&4.15298$\times10^{-07}$ ($<\alpha$)&2.93251$\times10^{-20}$ ($<\alpha$)\\
		\multicolumn{1}{ l }{Cliff's Delta effect size} & 1.0  $ (Large) $ & 0.973144 $ (Large) $ \\
		\bottomrule
	\end{tabular}
	\label{tab:div-results}
\end{table*}

\subsubsection{Results of Experiment on ArduCopter}
To answer RQ1, the results of the comparison between \aitester{} and Random testing based on violations of the expected behavior of the SUT (OCL violations) are presented in \Cref{ardu-results}. 
The table shows the total and the unique number of OCL violations corresponding to different UAV flight states. 
One constraint may be violated multiple times when the UAV is in one state for a particular time. 
For example, if a UAV takes five seconds to reach the altitude specified for the \textit{\mbox{Climb}} state and the environment changes are being observed after 500 milliseconds, one constraint for \textit{\mbox{Climb}} state can violate 10 times. 
Therefore, repeated violation of the same OCL constraint is considered as one unique violation. 
To isolate unique OCL violations, we first grouped recurring violations by flight state, followed by a manual analysis to ascertain the uniqueness of each OCL violation. 
We keep track of all violations as well as the unique number of violations. 
From the table, it can be observed that the number of OCL violations done by the \aitester{} in each flight state is greater than that of Random. 
It can be seen that the total number of OCL violations of \aitester{} is approximately thrice compared to Random. 
For some states such as \textit{\mbox{Descent}}, \textit{\mbox{Loiter}}, and \textit{\mbox{Landing}} the \aitester{} violates OCL constraints more than 50\% compared to Random. 
Moreover, for some states like \textit{\mbox{Climb}} and \textit{\mbox{AltitudeHold}}, the difference in the number of violations done by \aitester{} is higher than that of Random. 
The difference between \aitester{} and Random is noteworthy in the unique number of OCL violations. 
In some states (e.g.,  \textit{\mbox{Climb}} and \textit{\mbox{AltitudeHold}},), the \aitester{} violates slightly more OCL constraints compared to Random. 
For the important flight states such as \textit{\mbox{Descent}}, \textit{\mbox{Loiter}}, and \textit{\mbox{Landing}}, \aitester{} violates all state invariants specified for these states, whereas Random can only violate a few. 
The total number of unique OCL violations done by \aitester{} is more than 60\% compared to Random. 
The results suggest that \aitester{} effectively violated several OCL constraints representing the expected behavior of ArduCopter. 
These violations of OCL constraints may lead to faults in ArduCopter. 
However, identifying the root causes of faults in such a system requires faults reproduction and faults localization~\cite{wang2021exploratory} which is not in the scope of our approach.

\begin{acrqbox}
  \aitester{} is effective in generating diverse test scenarios that lead to deviations from the expected behavior of ArduCopter. 
\end{acrqbox}

To answer RQ2, the results of the comparison between \aitester{} and Random based on diversity are given in \Cref{tab:div-results}. 
The results of the Wilcoxon test suggest a statistically significant difference (i.e., \emph{p-value} $< \alpha$) between \aitester{} and Random. 
To analyze which one is better, the results of the Cliff's Delta effect size measure show a large effect size. 
This implies that the proposed approach (\aitester{}) is better in comparison to Random. 
Therefore, we can conclude that \aitester{} outperforms Random in exploring diverse flight paths.

\begin{acrqbox}
  \aitester{} is able to explore the diverse nature of flight paths from the behavioral model of UAV during testing. 
\end{acrqbox}

\begin{table*}[!t]
	\centering
	\noindent
	\caption{Comparison results from 100 episodes of \aitester{} and CDS Testing baseline approach for GCS-CDS case study}
	\begin{tabular}{l N N N N N N}\toprule
		\multicolumn{1}{l }{\textbf{}} & \multicolumn{3}{c }{\textbf{CDS Testing Approach}} & \multicolumn{3}{c }{\textbf{\aitester{} Approach}} \\ 
		\cmidrule(lr){2-4}
		\cmidrule(ll){5-7}
		\multicolumn{1}{ l }{\textbf{States}} & \textbf{Images} & \textbf{OCL Violations} & \textbf{Unique OCL Violations}& \textbf{Images} & \textbf{OCL Violations} & \textbf{Unique OCL Violations}\\ 
		\cmidrule(lr){1-1}
		\cmidrule(lr){2-4}
		\cmidrule(ll){5-7}
		\multicolumn{1}{ l }{\textbf{Armed}} & 441 & 30 & 1 & 171 & 60 & 3\\
		\multicolumn{1}{ l }{\textbf{AltitudeHold}} & 5744 & 1830 & 3 & 3107 & 3792 & 3\\
		\multicolumn{1}{ l }{\textbf{Approach}} & 866 & 838 & 2 & 55 & 92 & 3\\
		\multicolumn{1}{ l }{\textbf{Circle}} & 1902 & 2180 & 2 & 2521 & 3180 & 3\\
		\multicolumn{1}{ l }{\textbf{Climb}} & 7599 & 5550 & 3 & 6051 & 6644 & 3\\
		\multicolumn{1}{ l }{\textbf{Cruise}} & 5134 & 3620 & 3 & 3667 & 4078 & 3\\
		\multicolumn{1}{ l }{\textbf{Descent}} & 6405 & 3914 & 3 & 17819 & 20103 & 3\\
		\multicolumn{1}{ l }{\textbf{Landing}} & 963 & 960 & 3 & 62 & 61 & 3\\
		\multicolumn{1}{ l }{\textbf{Loiter}} & 1837 & 907 & 2 & 283 & 426 & 3\\
		\multicolumn{1}{ l }{\textbf{Takeoff}} & 788 & 405 & 2 & 63 & 11 & 3\\
		\cmidrule(ll){1-7}
		\multicolumn{1}{ l }{\textbf{\textit{Total}}} & \textit{31679} & \textit{20234} & \textbf{\textit{24}} & \textit{33799} & \textit{38447} & \textbf{\textit{30}}\\ 
		\bottomrule
	\end{tabular}
	\label{gcscds-results}
\end{table*}

\begin{figure*}
	\subfigure[A fault in ASI during Cruise state]{\includegraphics[width=3.1cm,height=3.1cm]{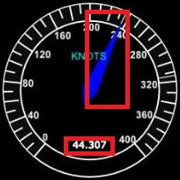}\label{bug1}}\hfill
	\subfigure[A fault in Altimeter during Approach state]{\includegraphics[width=3.1cm,height=3.1cm]{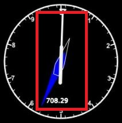}\label{bug2}}\hfill
	\subfigure[A fault in Altimeter during Approach state]{\includegraphics[width=3.1cm,height=3.1cm]{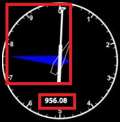}\label{bug3}}\hfill
	\subfigure[A fault in Altimeter during Approach state]{\includegraphics[width=3.1cm,height=3.1cm]{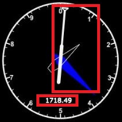}\label{bug4}}\hfill
	\subfigure[A fault in Altimeter during Approach state]{\includegraphics[width=3.1cm,height=3.1cm]{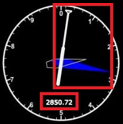}\label{bug5}}\hfill
	\subfigure[A fault in Turn Coordinator during Circle state]{\includegraphics[width=3.1cm,height=3.1cm]{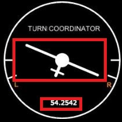}\label{bug6}}\hfill
	\subfigure[A fault in Turn Coordinator during Descent state]{\includegraphics[width=3.1cm,height=3.1cm]{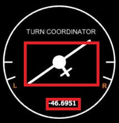}\label{bug7}}\hfill
	\subfigure[A fault in Turn Coordinator during Loiter state]{\includegraphics[width=3.1cm,height=3.1cm]{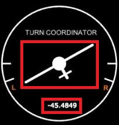}\label{bug8}}\hfill
	\caption{Faults identified in three CDS elements of GCS during different flight states}\label{gcscds-faults}
\end{figure*}

\subsubsection{Results of Experiment on GCS-CDS}
\Cref{gcscds-results} shows the results of the comparison between the \aitester{} approach and CDS testing baseline approach~\cite{sartajtesting}. 
\Cref{gcscds-faults} presents diagrams created from the results generated by \aitester{}. These diagrams depict the faults identified in three different flight instruments of GCS-CDS.

To answer RQ1, the results show that the total number of OCL violations performed by the \aitester{} is greater than that of the CDST approach. 
In most of the flight states, the \aitester{} performed a higher number of OCL violations compared to the CDST approach. 
The total number of unique OCL violations done by the \aitester{} is 30 whereas the CDST approach can perform 24 total unique OCL violations. 
The analysis of flight state-wise unique OCL violations shows that the number of violations by \aitester{} is greater than the CDST approach in some states, i.e., \textit{\mbox{Armed}}, \textit{\mbox{Approach}}, \textit{\mbox{Circle}}, \textit{\mbox{Loiter}}, and \textit{\mbox{Takeoff}}. 

The unique OCL violations performed by \aitester{} lead to four different types of faults (\Cref{gcscds-faults}) in GCS-CDS under test. 
These four types of faults are identified due to the violations of four OCL constraints shown in \Cref{gcscds-violated}. 
Due to the violation of OCL constraint \textit{C1}, the first type of fault is found in ASI during \textit{\mbox{Cruise}} flight state. 
According to the constraint \textit{C1}, the airspeed value must be between 50 and 100. 
\Cref{bug1} shows that the airspeed value pointed by the ASI needle is 240 knots whereas the ASI tape shows a value of $\approx$44 knots. 
This indicates an inconsistency between the ASI needle and tape. 
Therefore, this type of fault is considered a CDS elements integration fault.

\begin{lstlisting}[label=gcscds-violated, language=ocl, caption={Failed OCL constraints corresponding to four faults in GCS-CDS}, linewidth=13cm, numbers=none]
C1:context ArduCopter inv: self.oclIsInState(Cruise) and self.airspeedindicator.airspeed>=50 and self.airspeedindicator.airspeed<=100
C2:context ArduCopter inv: self.oclIsInState(Approach) and self.altimeter.altitude>=584 and self.altimeter.altitude<604
C3:context ArduCopter inv: self.oclIsInState(Circle) and (self.turncoordinator.roll>0 and self.turncoordinator.roll<=45) or (self.turncoordinator.roll<0 and self.turncoordinator.roll>=-45)
C4:context ArduCopter inv: self.oclIsInState(Loiter) and (self.turncoordinator.roll>0 and self.turncoordinator.roll<=45) or (self.turncoordinator.roll<0 and self.turncoordinator.roll>=-45)
\end{lstlisting}

Due to the violation of OCL constraint \textit{C2}, the second type of fault is found in the Altimeter during \textit{\mbox{Approach}} flight state. 
The constraint \textit{C2} states that the altitude value must be between 584 and 604 feet during the Approach state. 
At the same flight state and in different executions, the Altimeter has shown incorrect values. 
The Altimeter shows the altitude value of $\approx$ 709 feet in \Cref{bug2}, $\approx$ 956 feet in \Cref{bug3}, $\approx$ 1718 feet in \Cref{bug4}, and $\approx$ 2850 feet in \Cref{bug5}.

The third and fourth types of faults are identified in Turn Coordinator during \textit{\mbox{Circle}} and \textit{\mbox{Loiter}} states due to the violation of OCL constraints \textit{C3} and \textit{C4} respectively. 
In the third type of fault, Turn Coordinator shows a bank angle out of the range specified by OCL constraint \textit{C3}. 
Due to the out-of-range angle, the Turn Coordinator shows a high angle in \textit{\mbox{Circle}} state as shown in \Cref{bug6}. 
The fourth type of fault is identified due to the violation of OCL constraint \textit{C4}. 
In this fault, while turning left during \textit{\mbox{Descent}} (\Cref{bug7}) and \textit{\mbox{Loiter}} (\Cref{bug8}) states, the Turn Coordinator shows values higher than -46. 
It was analyzed that the UAV was not flipped upside down during both states. 
This happened as the angle exceeded the maximum limit during \textit{\mbox{Descent}} and \textit{\mbox{Loiter}} states.

All four types of faults identified by \aitester{} conform to the four types of faults found by the CDST approach~\cite{sartajtesting} in GCS-CDS. 
Therefore, the proposed approach (\aitester{}) is effective in finding potential faults in the GCS-CDS under test.

\begin{cdsrqbox}
  \aitester{} is effective in finding four potential faults in three different flight instruments of GCS-CDS. 
\end{cdsrqbox}

For the RQ2, the results of the comparison between the \aitester{} and CDST approach based on diversity are given in \Cref{tab:div-results}. 
The results of the Wilcoxon test suggest that \emph{p-value} $< \alpha$, i.e., there is a statistically significant difference between the \aitester{} and CDST approach. 
To identify which one is better, the results of the Cliff's Delta effect size measure show a large effect size. 
The results of statistical tests suggest that the proposed approach (\aitester{}) is better compared to the CDST approach. 
That is, \aitester{} outperforms the CDST approach in exploring the diverse nature of flight paths of UAV flight state machines.

\begin{cdsrqbox}
  \aitester{} is capable of exploring diverse flight paths from the UAV flight behavioral model during the testing of GCS-CDS.
\end{cdsrqbox}

\begin{figure}[!t]
	\centerline{\includegraphics[width=\linewidth,height=\textheight,keepaspectratio]{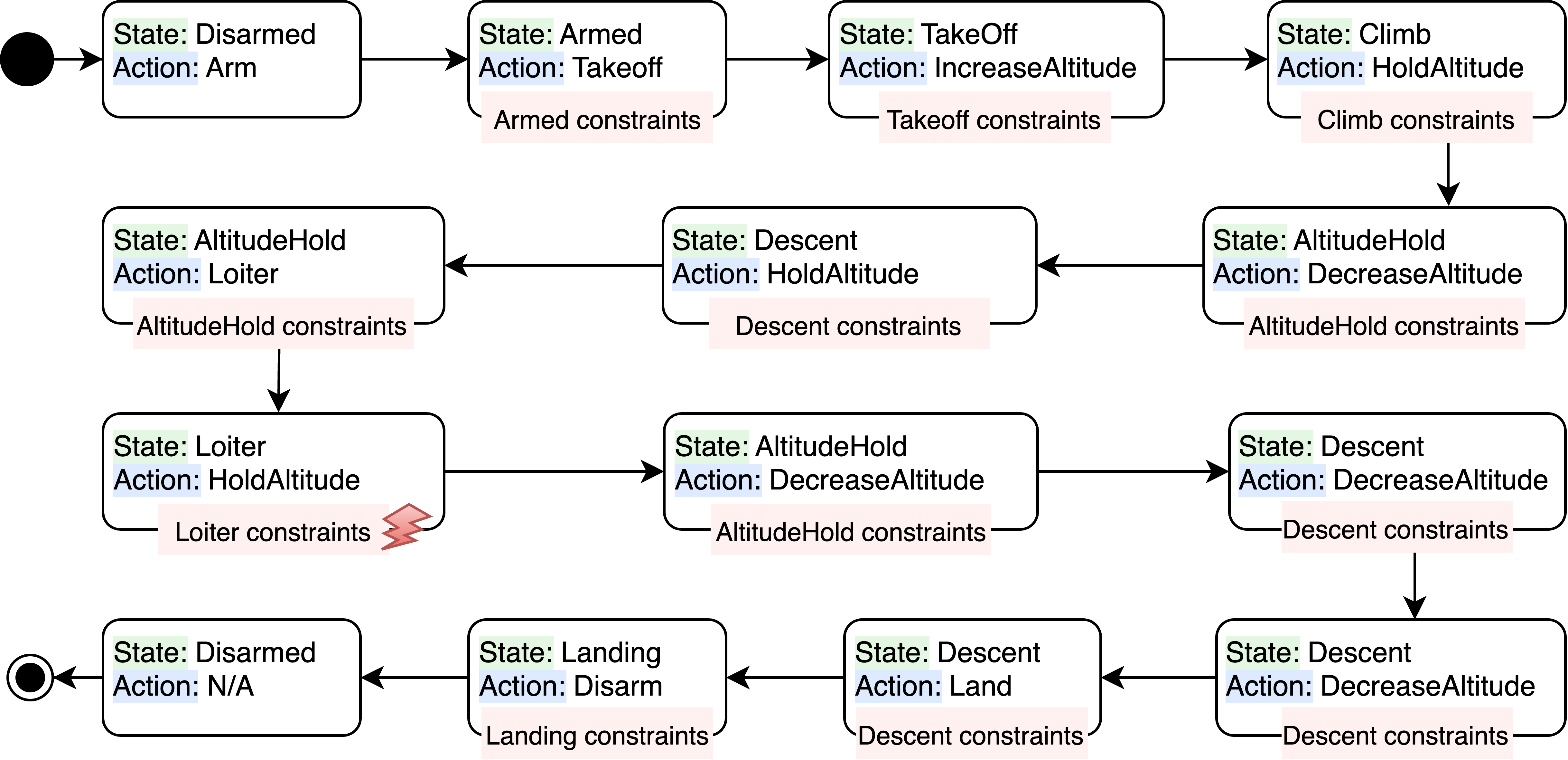}}
    \caption{A test path selected from experimental results, explored by \aitester{} during the testing process. It illustrates the sequence of states, actions, and evaluations of OCL constraints at each step along the path. This path eventually led to locating a fault in GCS-CDS shown in \Cref{bug8} during the \textit{\mbox{Loiter}} state due to the violation of an OCL constraint \textit{C4} (\Cref{gcscds-violated}).}
	\label{fig:testpath}
\end{figure}

\begin{figure}[!t]
    \centerline{\includegraphics[width=10.7cm,height=5.0cm,keepaspectratio]{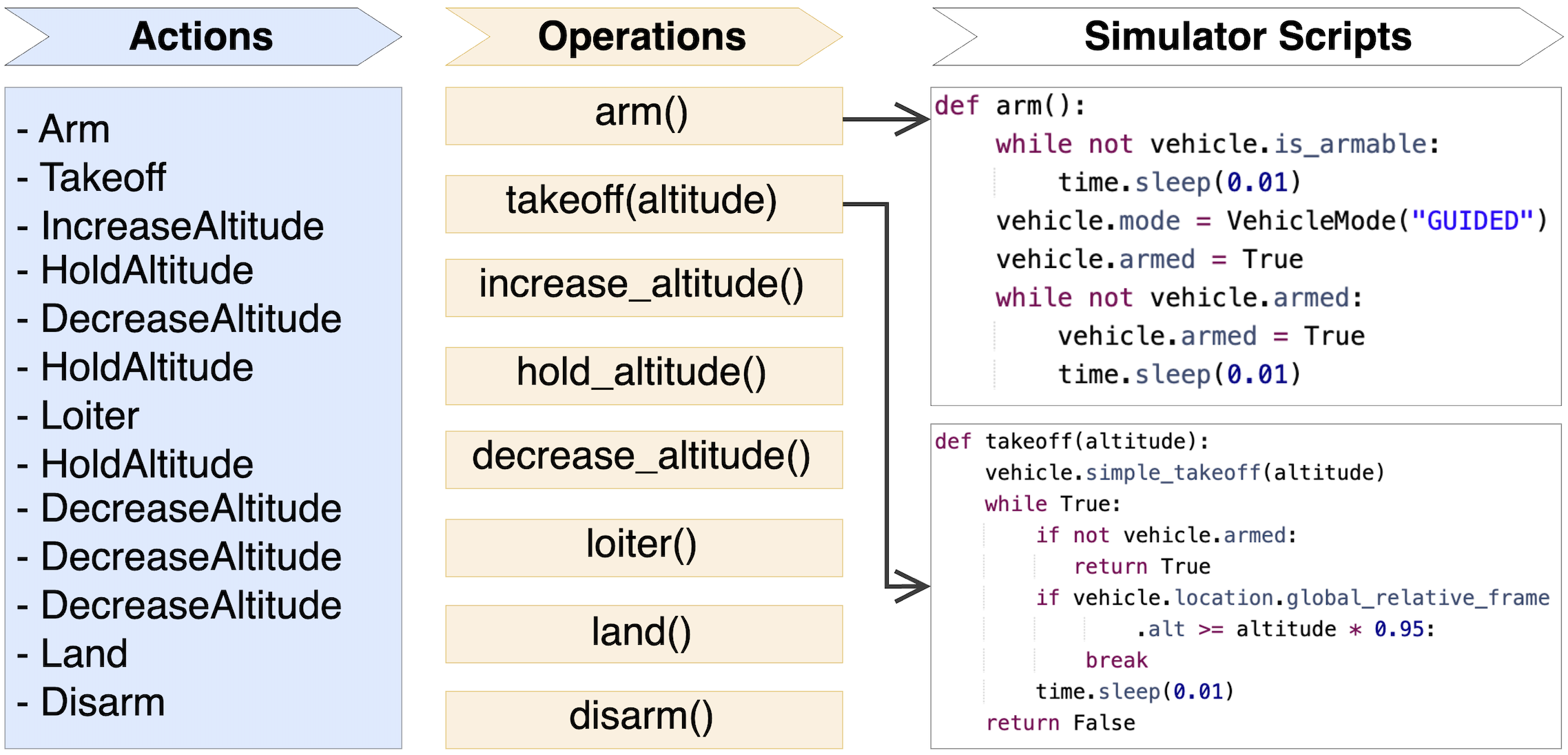}}
    \caption{Transformation of the test path from \Cref{fig:testpath} into an executable test script.}
	\label{fig:tpscript}
\end{figure}

\subsubsection{\textcolor{black}{Test Paths in Experiments}}
\textcolor{black}{
During the experiments, \aitester{} generated multiple test paths, with violations of OCL constraints occurring in various flight states. 
\Cref{fig:testpath} illustrates one of the test paths selected from the experimental results. 
This test path corresponds to a fault observed in the \emph{Turn Coordinator} of the GCS-CDS, as shown in \Cref{bug8}.
The fault occurred during the \textit{\mbox{Loiter}} state due to the violation of an OCL constraint \textit{C4} (\Cref{gcscds-violated}). 
While traversing this test path, \aitester{} executed the \emph{Arm} action when the UAV was in the \emph{Disarmed} state, resulting in a transition to the \emph{Armed} state. 
After the state change, \aitester{} evaluated the constraints specific to the \emph{Armed} state. 
Next, \aitester{} performed \emph{Takeoff} action which led to the \emph{TakeOff} state, and proceeded with the evaluation of constraints associated with this state. 
In a similar manner, \aitester{} selected the actions \emph{IncreaseAltitude}, \emph{HoldAltitude}, \emph{DecreaseAltitude}, \emph{HoldAltitude}, and \emph{Loiter}. 
These actions guided the UAV through the \emph{Climb}, \emph{AltitudeHold}, \emph{Descent}, \emph{AltitudeHold}, and \emph{Loiter} states, respectively. 
Simultaneously, \aitester{} evaluated the constraints applicable to each state. 
One of the constraints for \emph{Loiter} state was violated, indicating a deviation from the expected behavior. 
\aitester{} continued the testing process and performed further actions to take the UAV into \emph{AltitudeHold}, \emph{Descent}, \emph{Descent}, \emph{Landing}, and finally to \emph{Disarmed} states. 
Upon completing this process, the deviation reported by \aitester{} was utilized to locate the fault. 
}

\textcolor{black}{\textbf{Reproducing Test Paths.}
Each \aitester{}-generated test path consists of a series of actions executed by \aitester{} during the testing process. 
To reproduce the test paths generated by \aitester{}, each action must be transformed into corresponding UAV flight operations. 
\Cref{fig:tpscript} shows the transformation of a test path,  as depicted in \Cref{fig:testpath}, into an executable test script. 
Initially, for each action in the test path, a corresponding UAV flight operation call must be defined within the test script. 
For instance, \emph{Arm} action would be transformed into \emph{arm()} operation. 
Subsequently, for each UAV operation call, a script specific to the flight simulator must be defined. 
As illustrated in \Cref{fig:tpscript}, the scripts for \emph{arm()} and \emph{takeoff()} operations correspond to the Ardupilot flight simulator. 
If a different flight simulator is used, scripts for each operation need to be specifically defined for that particular simulator. 
Ultimately, this procedure results in test scripts corresponding to the test paths.
These scripts can be utilized to re-execute the test path. 
Note that the process of transforming test paths into test scripts can be automated using the CDST tookit~\cite{sartaj2020cdst}. 
}

\subsection{Threats to Validity}
The possible threats to the validity of our experiment are discussed individually in the following subsections. 

\subsubsection{External Validity Threat}
The possible threats to the validity of our experiment are discussed individually in the following subsections. 
The external threat to validity is related to the generalization of the experiment results. 
To reduce the chance of external validity threat, the experiment was performed using two primary subsystems of UAS, an autopilot (ArduCopter) and the CDS of GCS (GCS-CDS). 
ArduCopter is a good representative open-source autopilot system that is widely used in research and development~\cite{ma2019testing,koch2019reinforcement}. 
The ArduCopter supports a wide variety of multi-copter UAVs including a helicopter, tri-copter, quad-copter, hexacopter, etc. 
Moreover, this is also compatible with a wide range of hardware e.g., Pixhawk, CUAV v5 Plus, OpenPilot Revolution, etc. 
The GCS-CDS is an industrial case study containing widely used CDS elements. 
The results of our experiment may not be generalizable for all subsystems of UAS, however, this threat typically exists in every experiment~\cite{sartaj2024efficient}.

\subsubsection{Internal Validity Threat}
The parameter tuning of the training algorithm presents an internal threat to the validity of our experiment. 
There are no well-defined standard guidelines for selecting the hyperparameters of the DRL algorithm. 
We followed the common practice of selecting hyperparameters to minimize the internal validity threat. 
We performed several experimental trials with different values of parameters and their combinations. 
At the end of these trials, we selected the best-suited hyperparameters for our experiment. 
To reduce the chances of construct validity threat, we used the same stopping criterion for our approach (\aitester{}) and Random algorithm. 
We execute \aitester{} for 1000 episodes and Random for 1000 iterations during training. 
For the evaluation phase of both experiments, we repeated each approach 100 times which is a higher number compared to traditional DRL evaluation procedure~\cite{mnih2015human}. 
In the experiment with GCS-CDS, we configured and executed test paths according to the method defined in previous work~\cite{sartajtesting}.

\subsubsection{Construct Validity Threat}
Construct validity threat occurs when the relationship between cause and effect cannot be determined. 
To mitigate the possibility of this threat, we analyzed the experimental results according to the guidelines provided by Arcuri and Briand~\cite{arcuri2011practical}. 
We examined the experimental results based on their statistical significance and also analyzed the effect size. 
Specifically, we used the Wilcoxon signed-rank test for statistical significance and the Cliff's Delta effect size measure~\cite{cliff1993dominance} to analyze the effect size. 

\subsubsection{Conclusion Validity Threat}
The threat to conclusion validity is associated with the effect of treatments on the outcomes. 
To reduce the chances of this threat, we used the moving average reward (MAR) metric to analyze the reward during the training session. 
Our approach utilizes UAV flight state machines to explore different flight behaviors. 
To analyze the diverse exploration of UAV flight states, we used path diversity as a metric that was employed in a similar context and different domains, e.g., for Web~\cite{fard2013feedback}. 
We also used the unique number of OCL constraint violations as a metric to compare \aitester{} with the CDST approach. 
For a fair comparison, this metric is common between this work and the previous work~\cite{sartajtesting}.

\section{Related Works}\label{sec:rw}
This section presents a discussion of the works related to our approach including the works targeting UAV testing, autopilot system testing, the application of reinforcement learning to fly the UAV, and the studies related to the reinforcement learning-based testing of the UAV.

\subsection{UAV Testing}
In the literature related to the testing of unmanned aerial vehicles (UAV), 
Johnson and Fontaine~\cite{johnson2001use} developed a hardware and software-in-the-loop platform to carry out flight testing of small UAVs. 
For this purpose, test scenarios are required to be created manually, however, the test execution is automated. 
Our approach automates the whole testing process. 
Sorton and Hammaker~\cite{sorton2005simulated} manually created fifteen test scenarios for testing UAV in the hardware and software-in-the-loop simulation. 
Shore and Bodson~\cite{shore2005flight} manually developed different flight tests to evaluate the proposed algorithm for the control and identification of configuration parameters of the UAV.
Both the above-mentioned techniques use manually written and limited set of flight tests, however, our approach can automatically generate and execute flight test scenarios. 
Jung and Tsiotras~\cite{jung2007modeling} created a hardware-in-the-loop (HIL) simulation environment that can be used to develop and design flight tests for UAV autopilot. 
The proposed HIL simulation environment supports the manual creation of test scenarios, whereas our approach can automatically generate test scenarios.
Keith and Paul~\cite{sevcik2008testing} developed a testing environment to test UAV missions in a real-world context. 
Our approach does not target the testing of UAV missions.
How et al.~\cite{how2008real} developed a real-time test environment that can incorporate multiple autonomous (aerial and ground) vehicles. 
To use the test environment, the flight test scenarios need to be created manually. 

Mullins et al.~\cite{mullins2017automated} proposed an approach to generate challenging test scenarios for testing autonomous vehicles. 
The approach utilizes several manually designed test scenarios, executes in simulation, and searches for the critical test scenarios based on different boundary conditions. 
The main difference is that our approach can automatically generate and execute the test scenarios, and does not require base test cases to generate new test cases. 
Schulte and Spencer~\cite{schulte2018board} presented a model-based approach for UAVs that identifies faults, and sources of faults, and performs action accordingly during autonomous flight. 
To evaluate the proposed approach, a set of manually designed flight tests are used. 
Our approach targets automated system-level testing but does not perform onboard diagnostics during the UAV flight. 
Ma et al.~\cite{ma2018modeling} proposed an uncertainty-based modeling framework to test the self-healing part of cyber-physical systems. 
The proposed approach was used to test the self-healing behavior of the UAV, whereas our approach automates the system-level testing of the UAV.

Ulbig et al.~\cite{ulbig2019flight} proposed an approach and various unit and system-level test scenarios for the testing of the Terrain Awareness and Warning System of the UAV. 
Our approach automatically generates and executes test scenarios. 
Sarkar et al.~\cite{sarkar2019pie} proposed an approach and tool to automatically perform the flight testing of the UAV. 
The approach utilizes supervised learning to train a model that can perform different flight operations. 
For this purpose, the UAV flight data is captured using a camera, and the data filtering and labeling are carried out manually. 
Since the approach highly depends on the flight data, the inconsistency in data and labeling errors may lead to unpredictable flight testing. 
Moreover, the test oracle is also manual. 
In contrast, our approach does not rely on flight data and automates the whole testing process. 
De et al.~\cite{de2019design} manually designed various flight test scenarios for testing the vertical takeoff and landing behavior of the UAV carrying different payloads. 
Rizk et al.~\cite{rizk2019development} developed a GNSS-based emulator to facilitate UAV testing without flying. 
The emulator is required to be configured manually for testing the UAV. 
Wu et al.~\cite{wu2019testing} manually created various test scenarios for testing multi-rotor UAVs under different wind conditions and various flight modes (e.g., hover and flip). 
Yang et al.~\cite{yang2019small} manually developed test cases to test the acoustic behavior of the UAV. 
Scanavino et al.~\cite{scanavino2019new} created a test environment to evaluate the performance of the UAV under different climate conditions. 
To use the testing environment, the flight test scenarios need to be developed manually. 
Wang et al.~\cite{wang2021exploratory} performed an exploratory study to identify and classify bugs in the source code of ArduPilot and PX4 autopilot software. 
In recent work, Sorbo et al.~\cite{sorbo2022automated} conducted an empirical study to evaluate machine learning techniques for ensuring the safety of UAVs. 
A classification of potential sources of risks and accidents is presented based on the findings.
In comparison to the above-mentioned works, our approach can automatically perform the system-level testing of the UAS.

Several approaches are available targeting different aspects such as tests generation from flight logs~\cite{khatiri2023simulation}, analysis of boundary functions' role in UAVs safety~\cite{liang2021understanding}, faults identification using model checking~\cite{taylor2021avis}, detecting errors in the physical unit of UAS~\cite{taylor2022sa4u}, and mutation-based testing of robots~\cite{cavalcanti2019testing}. 
Furthermore, some studies are conducted focusing on characterizing UAVs software faults such as~\cite{sorbo2022automated,timperley2018crashing} and UAS autopilot faults~\cite{taylor2021study}.
In relation to these works, we focus on automating system-level testing of UAS.

\subsection{Autopilot Testing}
To test the autopilot of an aircraft, 
Fleck T.~\cite{fleck1983flight} manually designed various flight test scenarios considering the different ground and aerial situations such as the high speed at low-level, fly upwards, and altitude hold. 
Wells~\cite{wells1991tactical} presented various methods to test the flight control system of an autopilot. 
The proposed test methods target flight and missile stability control, and frequency response testing. 
Application and Flight Testing of an Adaptive Autopilot on Precision Guided Munitions
Sharma et al.~\cite{sharma2006application} manually developed nine flight tests to evaluate the performance of the proposed autopilot to support guided munitions. 
Kantue P.~\cite{kantue2017design} manually created various flight test scenarios to test the 
autopilot system developed for missile control. 
To test Piccolo II autopilot, Anderson et al.~\cite{anderson2008flight} performed different flight tests using the manually designed test scenarios to check different conditions such as GPS range and radio communication. 
Ambroziak and Gosiewski~\cite{ambroziak2013preliminary} manually defined various flight tests to analyze the variation in airspeed, altitude, and trajectories followed by the UAV.
Erdos and Watkins~\cite{erdos2008uav} manually created flight test scenarios to test the autopilot for the roll, pitch, altitude, airspeed, and navigation performance. 
Hartley et al.~\cite{hartley2013development} manually developed seven flight test scenarios to analyze the waypoint navigation of the autopilot. 
In all the above-mentioned works, the test scenarios are created manually. 
Whereas our approach automatically performs system-level testing of the UAS. 
He et al.~\cite{he2019system} proposed an approach to automatically generate test oracles for localizing faults in cyber-physical systems. 
The empirical evaluation using ArduPilot shows that the proposed approach generates better oracles compared to humans. 
Afzal et al.~\cite{afzal2021mithra} presented a technique to create oracles for cyber-physical systems based on historical telemetry data. 
The empirical evaluation using ArduPilot shows that the proposed technique outperforms the oracle generation approach by He et al.~\cite{he2019system}. 
In comparison, our work focuses on the online testing of UAS using OCL-based test oracles. 
Recently, Leach et al.~\cite{leach2022start} presented a framework for assessing and handling the security situation during UAV operation. 
The evaluation using ArduPilot and Red Team attacks shows that the proposed framework can increase the dependability of autonomous systems. 
Our work targets the system-level testing of UAS instead of the security testing of UAVs.

\subsection{Reinforcement Learning for UAV}
Reinforcement learning has been widely applied to autonomously control the flight of different types of UAVs. 
Since our approach applies reinforcement learning for the automated system-level testing of the UAV, we discuss a few works.
Bagnell and Schneider~\cite{bagnell2001autonomous} proposed a reinforcement learning-based algorithm for the robust control of autonomous helicopters. 
Kim et al.~\cite{kim2004autonomous} used reinforcement learning for training the algorithm to autonomously fly the helicopter and perform different flight maneuvers such as hover. 
Abbeel et al.~\cite{abbeel2007application} proposed an approach that finds the flight dynamics of the helicopter from the flight data and then uses reinforcement learning to learn optimal flight dynamics and autonomously fly the helicopter for various modes (e.g., flip).
Koch et al.~\cite{koch2019reinforcement} developed a reinforcement learning-based environment for training an agent to control the attitude of a quadcopter. 
In recent work, Yamagata et al.~\cite{yamagata2020falsification} applied deep reinforcement learning algorithms to temper cyber-physical systems based on robustness.

\subsection{Reinforcement Learning-based UAV Testing}
A small amount of work has been done in the area of RL-based UAV testing. 
Junell et al.~\cite{junell2016self} devised an approach using reinforcement learning to reduce the number of trials during flight testing of a quadrotor.
Siddiquee et al.~\cite{siddiquee2019flight} designed a vision-based reinforcement learning agent to test the guidance system of a quadcopter.
Ma et al.~\cite{ma2019testing} developed two algorithms based on reinforcement learning to test the self-healing part of cyber-physical systems under different uncertainties. 
The proposed algorithms were used to test the self-healing behavior of the ArduCopter. 
Ritchie Lee~\cite{lee2019adastress} presented Adastress, an approach for reinforcement learning-based stress testing of safety-critical systems. Subsequently, Moss et al.~\cite{moss2020adaptive} extended Adastress for the flight trajectory planning system, incorporating the aspect of continuous decision-making into the reinforcement learning approach. 
While these works primarily focus on stress testing, our work mainly aims to automate system-level testing of UAS. 
Yuning He~\cite{he2022system} introduced SYSAI, an approach that leverages AI techniques to automate various testing activities, such as property checking and test generation, for complex systems within the Aerospace domain. 
In comparison, our approach specifically targets UAS and employs reinforcement learning to automate system-level testing. 
In recent work, Zhang et al.~\cite{zhang2021figcps} proposed a reinforcement learning-based approach for generating failure-causing inputs. 
The empirical evaluation results show the success of the proposed approach in failing the well-tested CPS.  
The main difference between these works and our work is that our approach applies DRL to automate the system-level testing of the UAV.  

\section{Limitations}\label{sec:limit}

In this section, we discuss the potential limitations of our work and outline prospective avenues for subsequent research.

\subsection{Models Correctness}

Our work provides a comprehensive modeling methodology and guidelines for avionics testers to model the SUT. 
Avionics testers need to allocate time to prepare models and constraints. Furthermore, the correctness of these models and constraints must be verified by the testers, which is a common method~\cite{iqbal2019model,sartajtesting}.
Our approach operates under the assumption that the models and constraints developed by avionics testers are accurate. Ensuring their correctness is essential for gaining confidence in the testing outcomes.
The verification of models/constraints' correctness requires a specialized approach, however, this aspect falls outside the scope of this paper. 
Several existing methods available for this purpose, such as~\cite{balaban2013simplification,clariso2017smart,altoyan2023proving}, can be utilized to check the correctness of models and constraints.

\subsection{State Machine Learning}
 
Our approach requires avionics testers to model the behavior of SUT in the form of UML state machines. 
These state machines play a crucial role in test scenario generation and execution by \aitester{}. 
Therefore, they need to be carefully developed by avionics testers, which require manual effort. 
Alternatively, these state machines could potentially be learned during the testing process~\cite{PeledVY02}. 
This necessitates a future study to analyze the effectiveness of state machines learned automatically versus those crafted by testers.

\subsection{Complex Constraints}

The OCL constraints employed in our approach comprehensively encompass the requisites for conducting system testing of UAS.
Complex constraints such as performance and probabilistic constraints are typically associated with specific testing focuses like performance testing or uncertainty-based testing. 
However, this diverges from the primary focus of our work. 
Incorporating these types of testing focuses into our approach presents an intriguing direction for future research.

\subsection{Testing UAVs' Swarm}

The approach presented in this paper is a preliminary attempt to automate UAS testing. 
Thus, it is designed with a focus on a single subsystem of UAS as a SUT, such as a UAV or GCS at any given time. 
Our evaluation was also conducted considering testing UAV autopilot and GCS-CDS each as separate SUTs. 
Testing multiple UAS subsystems, such as a swarm of UAVs, is also a critical aspect of UAS testing. 
However, without empirical evidence, it remains uncertain whether our approach can be effectively applied to testing a swarm of UAVs or multiple GCSs. Further research is required to analyze the potential applicability of our approach to multiple UAS subsystems like UAVs' swarms.

\section{Conclusion}\label{sec:con}
Automated system-level testing of the UAS is required to meet the testing guidelines set by different international safety standards. 
The prevailing industrial practice and existing research works require manual effort in testing any of the UAS subsystems. 
Manual testing of a UAS subsystem is a tedious activity and only a handful of scenarios can be tested. 
In this paper, we proposed a novel approach to automate the system-level testing of the UAS. 
The proposed approach utilizes MBT and AI techniques to create an \aitester{} that generates, executes, and evaluates test scenarios based on the environmental context at runtime. 
We provided a toolset to support testing automation and to encourage further research and development. 
We performed two experiments to evaluate \aitester{} using two different subsystems of the UAS. 
The first experiment was performed using the ArduCopter autopilot system and the second experiment was performed using an industrial case study of GCS-CDS. 
The results demonstrated that the \aitester{} is effective in creating diverse test scenarios that lead to deviations from the expected behavior of the UAV autopilot and in revealing potential flaws in the GCS-CDS. 

The approach presented in this paper is a fundamental step towards automating system-level testing of UAS. 
Our approach generates and executes test scenarios on the fly to find deviations from the expected behavior of UAS SUT. 
While the test scenarios obtained through this process can assist avionics testers in identifying faults during the initial development phases of UAS, they are not designed to qualify for test certifications specified by international safety standards. 
Furthermore, test certification qualifications are dependent on the practices of the avionics industry, specifically the safety standards they adhere to.

In the future, we aim to concentrate on potential research directions, such as ensuring model correctness, learning state machines automatically, handling complex constraints, automating fault localization, and testing UAV swarms.

\backmatter

\section{Acknowledgments}
This research work is supported through a research grant titled \lq Establishment of National Center of Robotics and Automation (NCRA)\rq\space by Higher Education Commission (HEC) Pakistan.
We would like to acknowledge anonymous reviewers for their valuable suggestions, which greatly contributed to the significant improvement of this paper.

\begin{appendices}

\section{Abbreviations}\label{secAbbrev}
\Cref{tab:abbrev} provides a comprehensive list of all the abbreviations used in this paper, along with their full forms.

\begin{table}[ht]
	\centering
	\caption{List of abbreviations with their corresponding full forms}
    \label{tab:abbrev}
	\begin{tabular}{|l|l|}
		\hline
		\textbf{Abbreviation} &\textbf{Full form}\\
		\hline
		UAS&Unmanned Aerial System\\\hline
        UAV&Unmanned Aerial Vehicle\\\hline
		GCS&Ground Control Station\\\hline
        CDS&Cockpit Display Systems\\\hline
		CPS&Cyber–Physical System\\\hline
        HIL&Hardware-in-the-Loop\\\hline
		SIL&Software-in-the-Loop\\\hline
        PIL&Pilot-in-the-Loop\\\hline
        ISR&Intelligence, Surveillance and Reconnaissance \\\hline
		SUT&System Under Test \\\hline
        MBT&Model-Based Testing\\\hline
        UML&Unified Modeling Language\\\hline
        OCL&Object Constraint Language\\\hline
        EMF&Eclipse Modeling Framework\\\hline
        RL&Reinforcement Learning\\\hline
        DRL&Deep Reinforcement Learning\\\hline
        MDP&Markov Decision Process\\\hline
        ANN&Artificial Neural Network\\\hline
        RNN&Recurrent Neural Network\\\hline
        CNN&Convolutional Neural Network\\\hline
        DQN&Deep Q-Network\\\hline
        LSTM&Long Short-Term Memory\\\hline
        API&Application Programming Interface\\\hline
        ASI&Airspeed Indicator\\\hline
        ASL&Above Sea Level\\\hline
        AGL&Above Ground Level\\\hline
        MAR&Moving Average Reward\\\hline
 \end{tabular}
\end{table}




\end{appendices}

\bibliography{refs}

\end{document}